%% file: tau-decay-e+e--QCD25.tex
\documentclass[preprint,12pt]{elsarticle}
\makeatletter
\def\ps@pprintTitle{
	\let\@oddhead\@empty
	\let\@evenhead\@empty
	\def\@oddfoot{\reset@font\hfil\thepage\hfil}
	\let\@evenfoot\@oddfoot
}
\makeatother

 \biboptions{comma,sort&compress}

\usepackage[table]{xcolor}

\usepackage{graphicx}
\usepackage{amsmath}
\usepackage{here}
\usepackage{cuted}

\usepackage{here}
\usepackage[dvips]{epsfig}
\definecolor{bluette}{rgb}{.2,.4,0}
\definecolor{salmon}{rgb}{.9,0.68,0.5}
\definecolor{motive}{rgb}{0.2,1,.5}
\definecolor{list}{rgb}{0.3,.8,.1}
\definecolor{moe}{rgb}{1,.7,.5}
\definecolor{mote}{rgb}{.7,.5,.6}
\definecolor{pisello}{rgb}{.1,1,0}
\definecolor{orange}{rgb}{1,.7,0}
\definecolor{oliva}{rgb}{.1,.5,0.3}
\definecolor{greenda}{rgb}{0,.3,.2}
\definecolor{greenli}{rgb}{0.5,.8,.0}
\definecolor{blueda}{rgb}{0,.1,.6}
\definecolor{purple}{rgb}{.7,.1,.2}
\definecolor{marrone}{rgb}{1,0.7,0}
\definecolor{pinky}{rgb}{1,0.8,0.8}
\definecolor{rose}{rgb}{1,0.4,0}

\def\oliva{\color{oliva}}

\def\beq{\begin{equation}}
\def\eeq{\end{equation}}
\def\bea{\begin{eqnarray}}
\def\eea{\end{eqnarray}}
\def\bq{\begin{quote}}
\def\eq{\end{quote}}

\def\nnb{\nonumber}
\def\ga{\left(}
\def\dr{\right)}

\def\lrar{\Longrightarrow}
\def\lrar2{\longrightarrow}
																
\def\nnb{\nonumber}

\def\la{\langle}
\def\ra{\rangle}

\def\ba{\vspace*{-0.2cm}\begin{array}}
\def\ea{\end{array}\vspace*{-0.2cm}}

\def\b{$\bullet~$}
\def\d{$\diamond~$}

\def\als{\alpha_s}

\def\gg2{\la\alpha_s G^2 \ra}
\def\gg3{g^3f_{abc}\la G^aG^bG^c \ra}
\def\ggg4{\la\als^2G^4\ra}

\def\gg{\lag g^{2}_{s} G^2 \rag}
\def\ggg{\lag g^{3}_{s}G^3\rag}

\usepackage{amsmath}
\usepackage{slashed}
\usepackage{color}

\begin{document}
\begin{frontmatter}

\title{QCD condensates and $\alpha_s$  from $e^+e^-$ and $\tau$-decays\,\tnoteref{invit}}
\tnotetext[invit]{{Review talk given at QCD25 (40th anniversary of the QCD-Montpellier Conference)}, 30th June - 4th July 2025, Montpellier-FR.}
\author{Stephan Narison
}
\address{Laboratoire
Univers et Particules de Montpellier (LUPM), CNRS-IN2P3, \\
Case 070, Place Eug\`ene
Bataillon, 34095 - Montpellier, France\\
and\\
Institute of High-Energy Physics of Madagascar (iHEPMAD)\\
University of Ankatso, Antananarivo 101, Madagascar}
\ead{snarison@yahoo.fr}


\date{\today}
\begin{abstract}
In this talk, I review the determinations of the 
 QCD condensates and $\alpha_s$ within the SVZ expansion using the ratio of Laplace sum rule (LSR)  and $\tau$-like moments in $e^+e^-\to I=1$ Hadrons and in $\tau\to \nu_\tau+$Hadrons  Axial-vector (A) and V-A channels. Some misprints in the original papers\,\cite{SNe,SNe2,SNtau24} have been corrected.  We found that the value of the gluon condensate  agrees with the one $\la \alpha_s G^2\ra=(6.35\pm 0.35)\times 10^{-2}$ GeV$^4$ from quarkonia and some other sum rules but less accurate, while the factorization of the four-quark condensate is violated by a factor about 6: $\rho\la\bar\psi\psi\ra^2=(5.98\pm 0.64)\times 10^{-4}$ GeV$^6$ which confirms previous findings. Extracting the QCD condensates up to dimension D=20 from $e^+e^-$ and the axial-vector channel of $\tau$-decay, we do not find  any exponential growth of their values in the Euclidian region,  thus excluding (by duality) any significant effect of the so-called Duality Violation (DV) in the Time-like one.  The optimal values of $\alpha_s(M_\tau)$ from  $e^+e^-\to$ Hadrons and $\tau$-decays agree each others and lead to the average:  $\alpha_s(M_\tau)=0.3128(51)$\, [resp.0.3330(57)] $\lrar2 \alpha_s(M_Z) = 0.1176$\, [resp. 0.1201] $(7)_{fit}(3)_{evol.}$ for Fixed Order (FO) [resp. Contour improved (CI)] PT series to be compared with the PDG24 average (without Lattice calculations)\,: $ \alpha_s(M_Z) = 0.1175(10)$. 
  \end{abstract}

\end{frontmatter}

\newpage
\section{Introduction}
\vspace*{-0.2cm}

A precise determination of the QCD  condensates is an important step for understanding the theory of QCD and its vacuum structure while the one of $\alpha_s$ is useful for PT QCD and as input in the precision test of the Standard Model (SM). 
In this paper,  I summarize the determinations of the QCD condensates  in Ref\,\cite{SNe,SNe2} from $e^+e^-\to$ I=1 Hadrons and in Ref.\,\cite{SNtau24} from the Axial-Vector and V-A channels of the semi-hadronic $\tau$-decays. I also correct some misprints in these original papers and improve the discussions of the obtained results.

\section{The two-point functions}
We shall be concerned with the two-point correlator :
\bea
\hspace*{-0.6cm} 
\Pi^{\mu\nu}_{V(A)}(q^2)&=&i\hspace*{-0.1cm}\int \hspace*{-0.15cm}d^4x ~e^{-iqx}\la 0\vert {\cal T} {J^\mu_{V(A)}}(x)\ga {J^\nu_{V(A)}}(0)\dr^\dagger \vert 0\ra, \nnb\\
&=&-(g^{\mu\nu}q^2-q^\mu q^\nu)\Pi^{(1)}_{V(A)}(q^2)+q^\mu q^\nu \Pi^{(0)}_{V(A)}(q^2),
 \label{eq:2-point}
 \eea
built from the T-product of the bilinear vector(V), axial-vector  (A) and V--A currents of $u,d$ quark fields with:
\bea
J^\mu_V(x)&=& \frac{1}{2}: \bar\psi_u\gamma^\mu\psi_u-\bar\psi_d\gamma^\mu\psi_d:\,,   \nnb\\
 J^\mu_{A}(x)&=&: \bar\psi_u\gamma^\mu\gamma_5\psi_d:,~~~~~
 J^\mu_{V-A}(x)=: \bar\psi_u\gamma^\mu(1-\gamma_5)\psi_d:.
\eea
The vector current corresponds to the $e^+e^-\to$I=1 Hadrons process while the axial-vector and V-A currents are from $\tau$-decay channels. 
The upper  indices (0) and (1) correspond to the spin of the associated hadrons. The two-point function obeys the dispersion relation:
\beq
\Pi_{V(A)}(q^2)=\int_{t>}^\infty \frac{dt}{t-q^2-i\epsilon} \frac{1}{\pi}\,{\rm Im} \Pi_{V(A)}(t)+\cdots,
\eeq
where $\cdots$ are subtraction constants polynomial in $q^2$ and $t>$   is the hadronic threshold.
\subsection*{\b  The QCD two-point function within the SVZ-expansion}
According to\,\cite{SVZ}, the two-point function can be expressed in terms of the sum of higher and higher quark and gluon condensates:
\beq
4\pi^2\Pi_H(-Q^2,m_q^2,\mu)=\sum_{D=0,2,4,..}\hspace*{-0.25cm}\frac{C_{D,H}(Q^2,m_q^2,\mu)\la O_{D,H}(\mu)\ra}{(Q^2)^{D/2}}\equiv \sum_{D=0,2,4,..}\hspace*{-0.25cm}\frac {d_{D,H}} {(Q^2)^{D/2}}~, 
\label{eq:ope}
\eeq
where $H\equiv V,A$ and $V-A$, $\mu$ is the subtraction scale which separates the long (condensates) and short (Wilson coefficients) distance dynamics and $m_q$ is the quark mass. 

\d The perturbative expression of the spectral function is known to order $\alpha_s^4$ as given explicitly in\,\cite{SNtau24,SNB2}. 
We shall use the value $\Lambda =(342\pm 8)$ MeV  for  $n_f=3$\:\, from the PDG average\,\cite{PDG}.

\d The non-perturbative contributions to the two-point function is given up to $d_{8,A}$ 
\,\cite{BNP,BNP2}.

\subsection*{\b The Vector (V)  spectral function}
\begin{figure}[hbt]
\begin{center}
\includegraphics[width=11.cm]{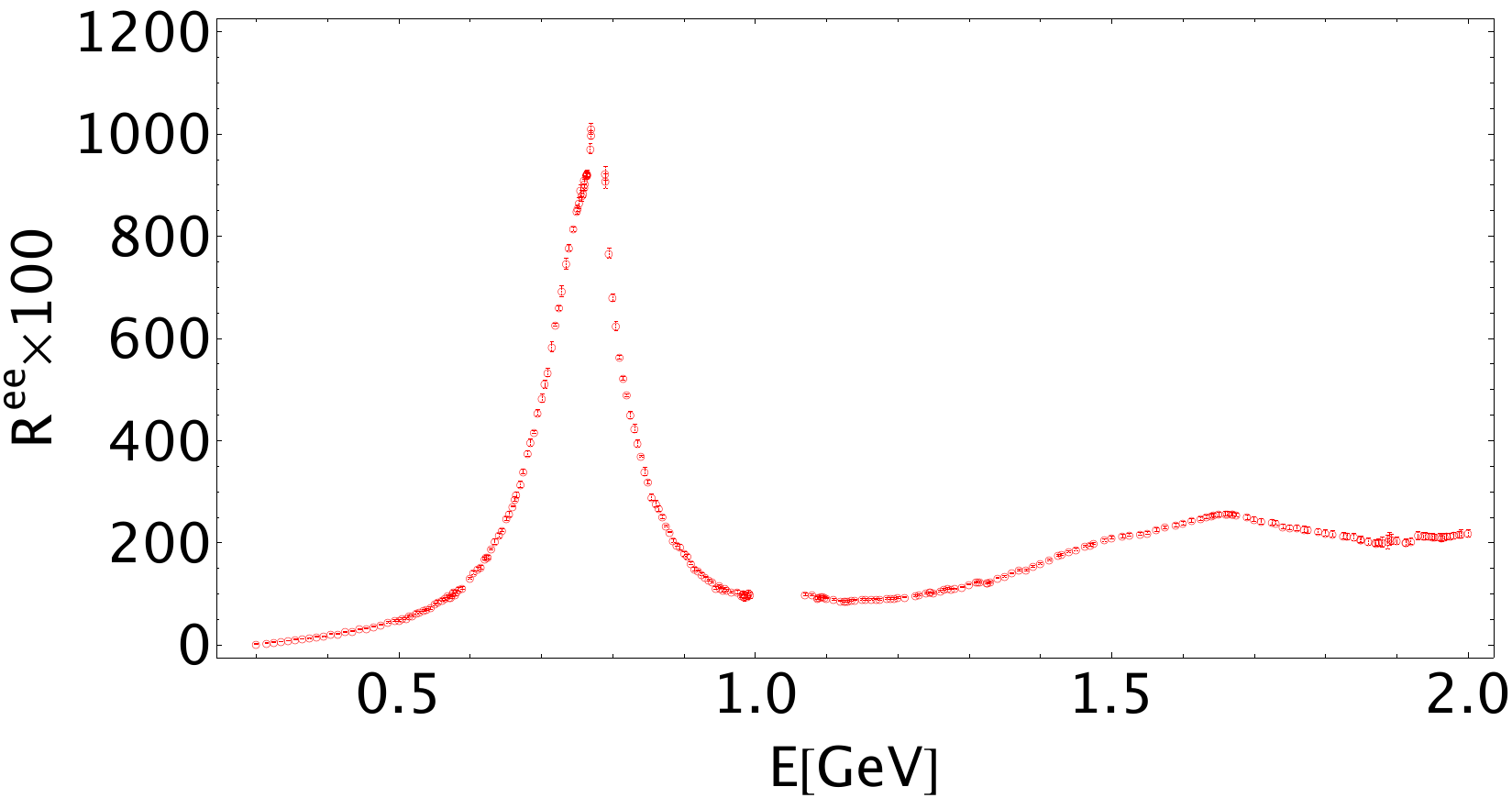}
\caption{\footnotesize  
The ratio $R$ of $e^+e^-\to I=1\,{\rm Hadrons}$ data as a function of the c.m. energy E. } \label{fig:sigma1}
\end{center}
\vspace*{-0.5cm}
\end{figure}

It corresponds to the $e^+e^-\to I=1$ Hadrons total cross-section. We shall use below 1 GeV the new data from CMD-3\,\cite{CMD3}
\footnote{Notice that unlike the case of the muon anomalous magnetic moment $a_\mu$, the result is about the same if one uses the PDG22 compilation below 1 GeV rarher than the CMD-3 data.}  and above 1 GeV the compilation of PDG22\,\cite{PDG}. 
We show the data of:
\beq
R^{ee}\equiv \frac{\sigma (e^+e^-\to I=1\,{\rm Hadrons})}{\sigma (e^+e^-\to \mu^+\mu^-)} ==\ga\frac{3}{2}\dr 
8\pi\, {\rm Im} \Pi_H(t).
\eeq
in Fig\,\ref{fig:sigma1}, which we fit with a simple Breit-Wigner (BW) fit below 1 GeV ($\rho$ meson), a polynomial  from 1 to  1.5 GeV  and a BW ($\rho'$) from 1.5 to 1.9 GeV. The separation between the I=1 and I=0 components are discussed in details in Ref.\,\cite{SNe}. 
We subdivide the regions into different subregions and fit the data using polynomials  with the Mathematice FindFit optimized program.
To be conservative, we fit separately the upper and lower parts of the data and consider as a final value their average. 
\subsection*{\b The Axial-vector (A)  and V--A spectral functions}
\begin{figure}[H]
\begin{center}
\includegraphics[width=7.3cm]{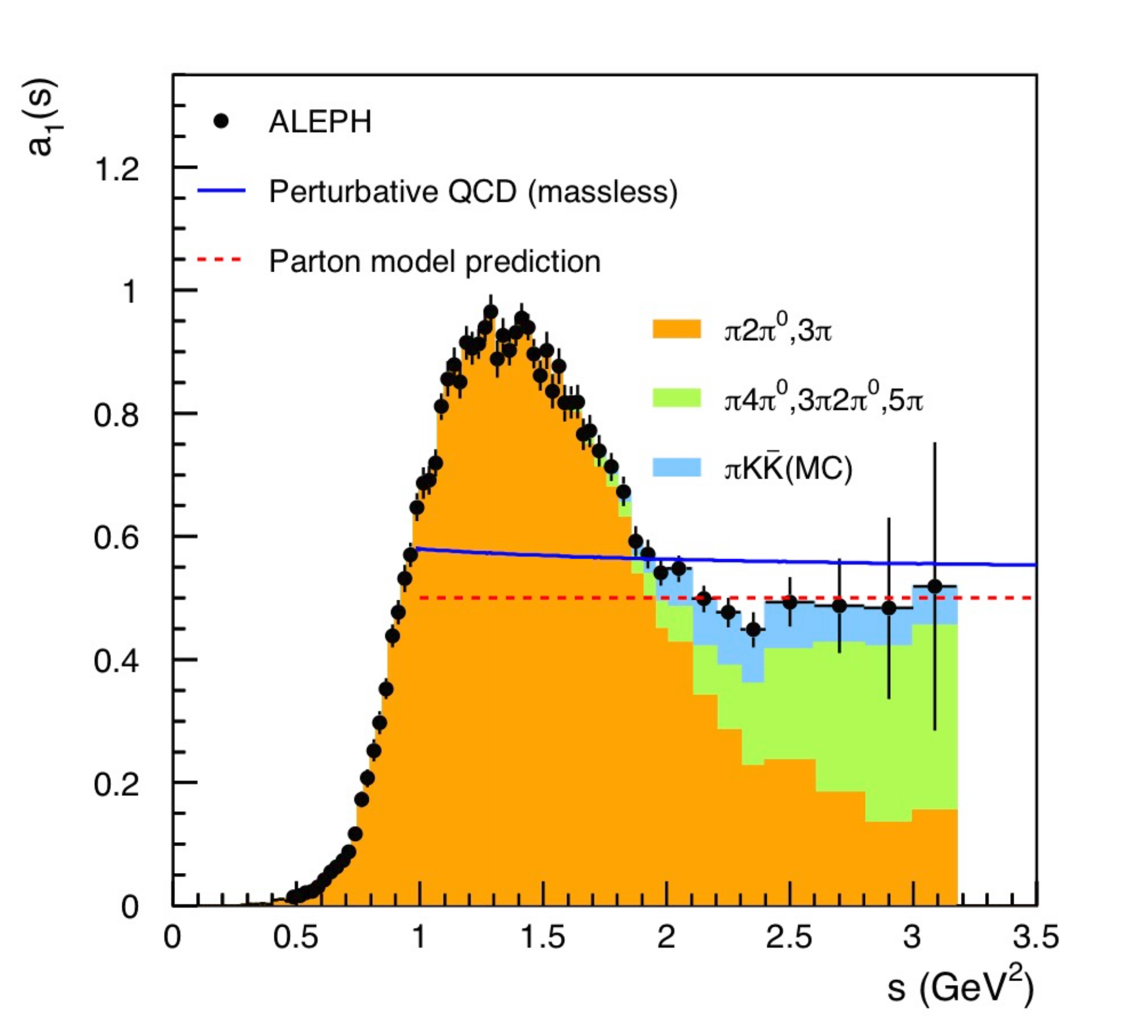}
\includegraphics[width=7cm]{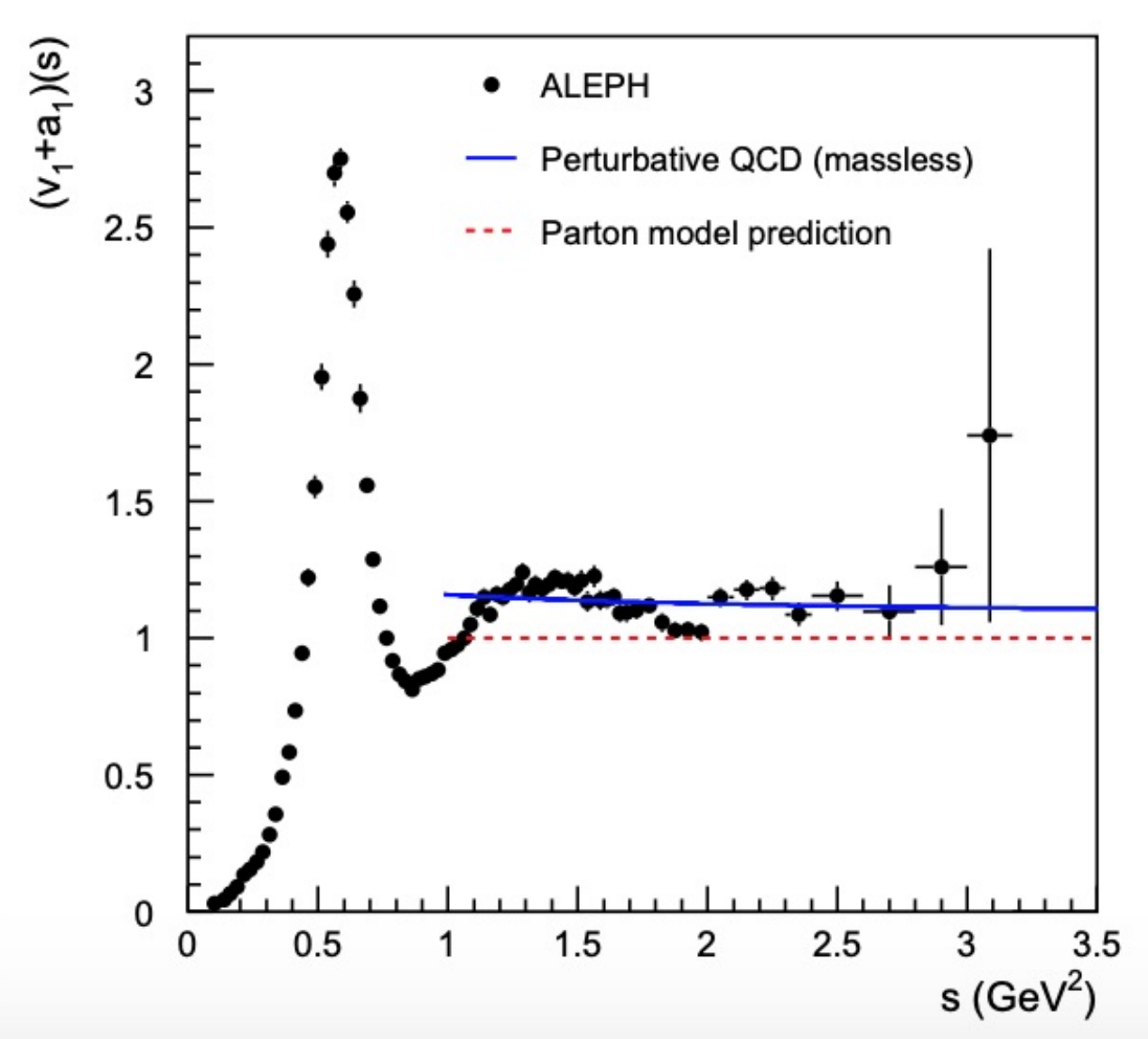}
\caption{\footnotesize  ALEPH data of the axial-vector and V--A spectral functions. } \label{fig:aleph}
\end{center}
\vspace*{-0.5cm}
\end{figure}

We shall use the recent ALEPH data\,\cite{ALEPH}  in Fig.\,\ref{fig:aleph} for the spectral function $a_1(s)$ and $v_1+a_1$ associated respectively to the A and V--A currents. 

Like in the case of the vector spectral function from $e^+e^-\to$ Hadrons\,\cite{SNe2}, we subdivide the region from threshold to the physical $\tau$ mass squared $M_\tau^2=3.16$ GeV$^2$ into different subregions 
and fit the data with 2nd and 3rd order polynomials using the optimized Mathematica program FindFit.  The details of the analysis are described in Ref.\,\cite{SNtau24}. 
\section{The ratio of Laplace sum rule (LSR) moments}
\d We shall work here with the ratio of LSR moments\,\cite{SVZ,SNR,BELL}\footnote{For a recent review, see e.g.\,\cite{SNLSR}.}  as in Ref.\,\cite{LNT}:
\beq
 {\cal R}^H_{10}(\tau)\equiv\frac{{\cal L}^c_{1}} {{\cal L}^c_0}= \frac{\int_{t>}^{t_c}dt~e^{-t\tau}t\, \frac{1}{\pi}\,{\rm Im}
 \Pi_H(t,\mu^2,m_q^2)}   {\int_{t>}^{t_c}dt~e^{-t\tau} \, \frac{1}{\pi}\,{\rm Im}
 \Pi_H(t,\mu^2,m_q^2)},
\label{eq:lsr}
\eeq
where  $\tau$ is the LSR variable, $t>$   is the hadronic threshold.  Here $t_c$ is  the threshold of the ``QCD continuum" which parametrizes, from the discontinuity of the Feynman diagrams, the spectral function  ${\rm Im}\,\Pi_H(t,m_q^2,\mu^2)$.  $m_q$ is the quark mass and $\mu$ is an arbitrary subtraction point.

\d The PT expression of the two-point function is known to order $\alpha_s^4$  as explicitly given in Ref.\,\cite{SNtau24} from which one can deduce the PT expression of the LSR moment:
\beq
{\cal L}^{PT}_0(\tau)= \frac{3}{2}\tau^{-1}\Big{[} 1+a_s+2.93856\,a_s^2+ 6.2985\,a_s^3 + 22.2233\,a_s^4\Big{]}.
\eeq

\d Within the SVZ-expansion, the non-perturbative 
contributions to the lowest LSR moment in terms of the $d_D$ condensates is\,:
\beq
{\cal L}^{NPT}_0(\tau) = \frac{3}{2}\tau^{-1}\sum_{D\geq 2}\frac{d_D}{(D/2-1)!} \tau^{D/2} ~,
\label{eq:svz}
\eeq
from which one can deduce ${\cal L}^{NPT}_1$ and ${\cal R}^H_{10}$. The expressions of $d_D$ including $\alpha_s$ corrections are given explicitly in Refs.\,\cite{SNe,SNtau24} and the original references given there. 

\section{The generalized  $\tau$-decay moments}
As emphasized by BNP in Ref.\,\cite{BNP,BNP2}, it is more convenient to express the moment in terms of the combination of Spin (1+0) and Spin 0 spectral functions in order to avoid some eventual pole from $\Pi^{(0)}$ at $s=0$.  
  \vspace*{-0.25cm}
\begin{figure}[H]
\begin{center}
\includegraphics[width=8cm]{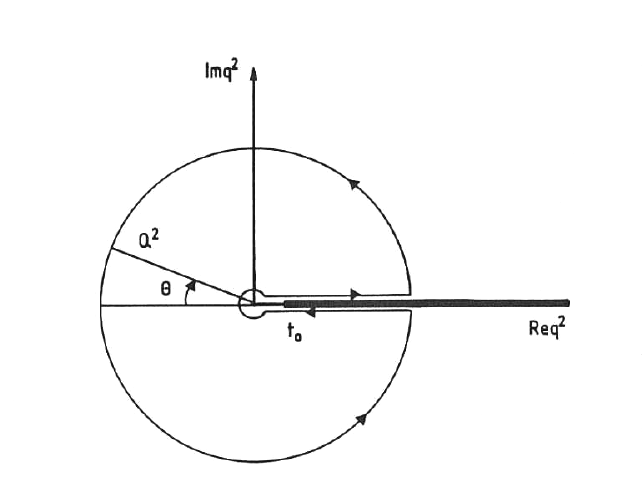}
\caption{\footnotesize Cauchy contour integral in the complex $Q^2\equiv |q^2|$-plane}\label{fig:cauchy}
\end{center}
\vspace*{-0.25cm}
\end{figure}  
Using the Cauchy contour in Fig.\,\ref{fig:cauchy}  \`a la Shankar\,\cite{SHANKAR}, the moment can be written as\,: 
  \beq
  {\cal R}_{n,H}=6\pi\,i\,\int_{|s|=M_0^2} \hspace*{-0.5cm}dx\, (1-x)^2\,x^n\ga (1+2x) \,\Pi^{(1+0)}_H(x) -2x\Pi^{(0)}_H(x)\dr,
  \eeq
  with $x\equiv s/M^2_0$, $H\equiv V,A, V-A$. $n$ indicates the degree of moment.  Some other variants of the moments have been presented in Ref.\,\cite{LEDI}. 
\subsection*{\b The lowest BNP moment ${\cal R}_{0,H}$}  
It corresponds to the physical $\tau$-decay process\,\cite{BNP2,BNP} .
Its QCD expression to order $\alpha_s^4$ and up to $d_8$ condensates
  is explicitly given in Ref.\,\cite{BNP2,BNP}.  The size of the  PT $a_s^5$ coefficient has been estimated by observing that the calculated coefficients of the  PT series grow geometrically\,\cite{SNZ}.  In the following, we consider this contribution as a source of systematic errors for the truncation of the PT series.
\subsection*{\b The higher $\tau$-decay moment ${\cal R}_{n,H}$}  
In the following, we shall work with the moments of degree up to $n=6$. Their QCD expressions up to order $\alpha_s^4$ are given in Ref.\,\cite{SNe2}. From these expressions, one can see that, to leading order in $\alpha_s$, the Cauchy theorem selects the dimension of the QCD condensates entering in the QCD expressions of the moments which enable to extract these condensates with a good confidence. 

\section{QCD condensates from the LSR ratio $R^{ee}_{10}$ using $e^+e^-$ data}

\d Truncating the OPE at $D=6$, we attempt to extract $(d_4,d_6)$ using a two-parameter fit. The analysis is not conclusive as we do not have $\tau$-stability.

\begin{figure}[hbt]
\begin{center}
\hspace*{-1cm}{\bf a)} \hspace*{7cm}{\bf b)}\\
\includegraphics[width=7.5cm]{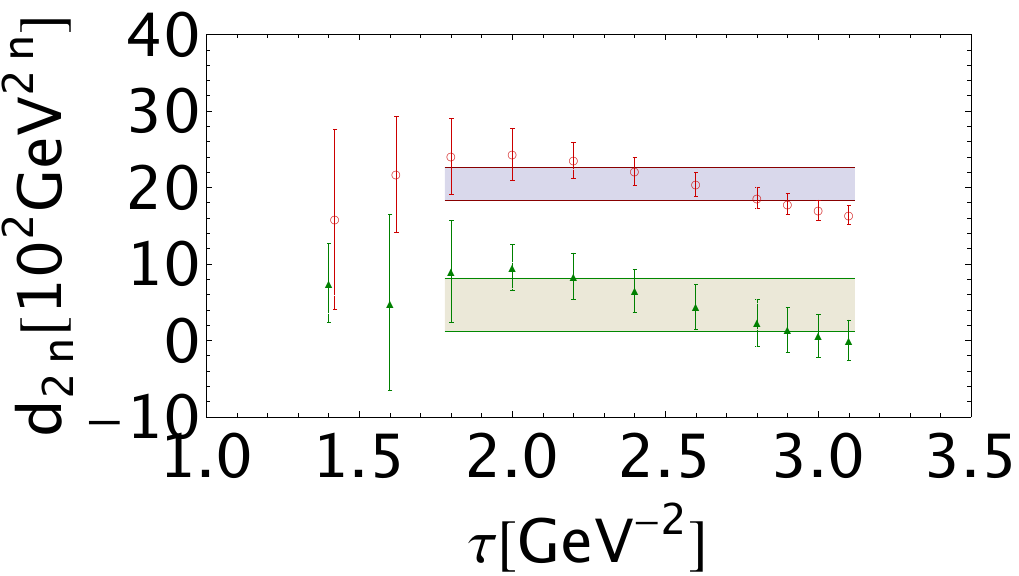}
\includegraphics[width=7.5cm]{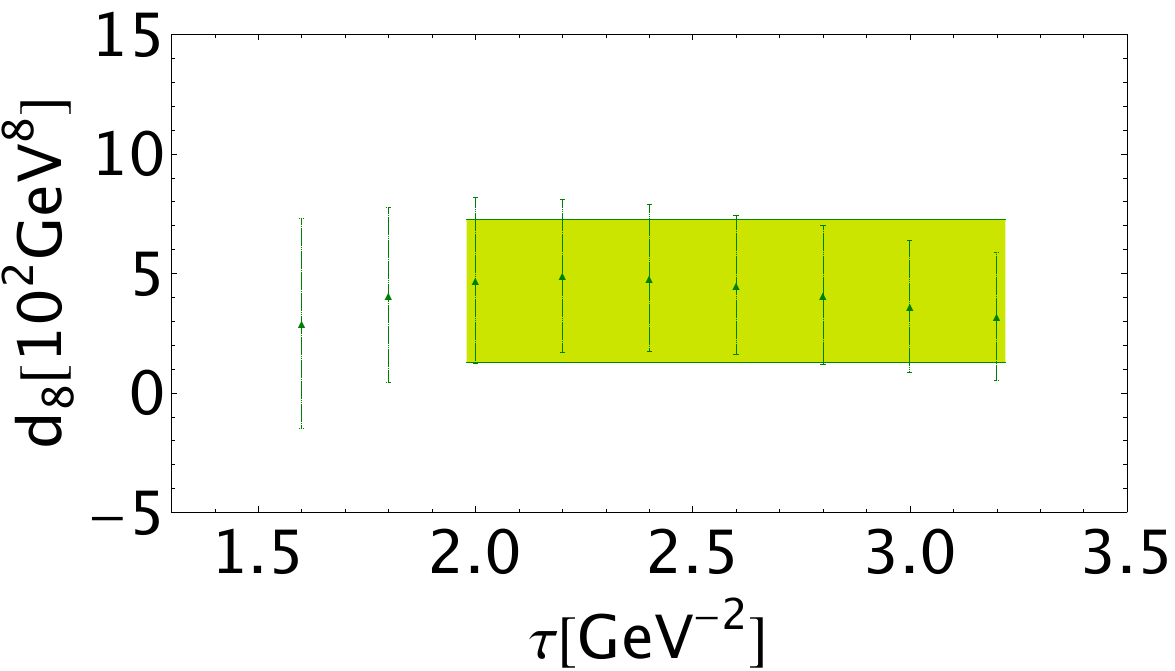}
\caption{\footnotesize  a) Two-parameter fit of $-d_6$ and $+d_8$ to order $\alpha_s^4$  for a given value of $\la\alpha_s G^2\ra$ from Eq.\,\ref{eq:asg2}; b) One-parameter fit of $d_8$ to order $\alpha_s^4$  for a given value of $\la\alpha_s G^2\ra$ from Eq.\,\ref{eq:asg2} and $d_6$ from Eq.\,\ref{eq:res-d68}.} 
\label{fig:d8-as4}.
\end{center}
\vspace*{-0.5cm}
\end{figure}

\d Then, we use as input the value of 
the gluon condesate from the heavy quark mass-splittings and some other  sum rules\,\cite{SNparam, SNcb1}:
\beq
\la\alpha_s G^2\ra =  (6.39\pm 0.35)\times 10^{-2}\,{\rm GeV^4}. 
\label{eq:asg2}
\eeq
We truncate the OPE up to $D=8$ and extract $d_6$ and $d_{8}$.  We show in Fig.\,\ref{fig:d8-as4}a the results from a two-parameter fit to order $\alpha_s^4$.  A least square fit of the points between 1.8 and 3.2 GeV$^{-2}$, where the determinations are more precise, gives:
\beq
d_6 =  -(20.5\pm2.2)\times 10^{-2}\,{\rm GeV^6},\,\,\,\,\,\,\, d_8= (4.7\pm 3.5)\times 10^{-2}\,{\rm GeV^8},
\label{eq:res-d68}
\eeq 
 where, one should note that the value of $d_8$ 
is largely affected by the truncation of the PT series.  The value of $d_6$ corresponds to:
\beq
 \rho\la\bar\psi\psi\ra^2=(5.98\pm 0.64)\times 10^{-4}\,{\rm GeV^6},
 \label{eq:res-4q}
 \eeq
 indicating a violation of about a factor 5.8 of the vacuum saturation assumption. 
 
\d To improve its value, we redo the fit by using as input $\la \alpha_s G^2\ra$ and the previous value of $d_6$. One deduces a much better stability in Fig.\,\ref{fig:d8-as4}b which gives the value:
\beq
d_8 = (4.3\pm 3.0)\times 10^{-2}\,{\rm GeV^8},
\label{eq:d8-res}
\eeq
which confirms the result in Eq.\,\ref{eq:res-d68}. 

\d One should note that the optimal value is obtained for a relatively large value of $\tau$ from 1.8 to 3.2 GeV$^{-2}$ and is uncomfortable. We shall come back to this estimate after the use of the $\tau$-like moments.

\begin{center}
\begin{figure}[hbt]
\begin{center}
\hspace*{-1cm}{\bf a)} \hspace*{7cm}{\bf b)}\\
\includegraphics[width=7.5cm]{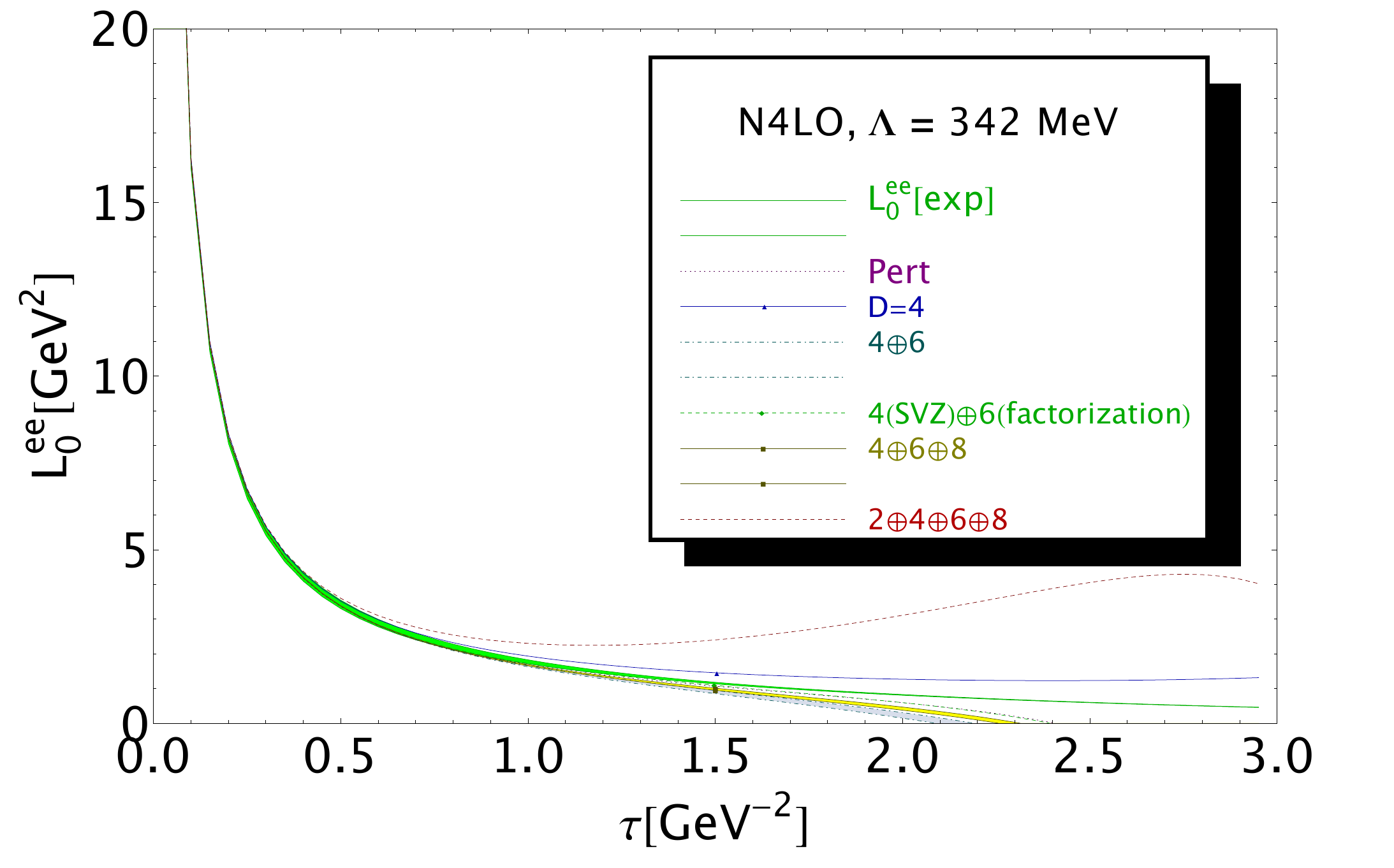}
\includegraphics[width=7.5cm]{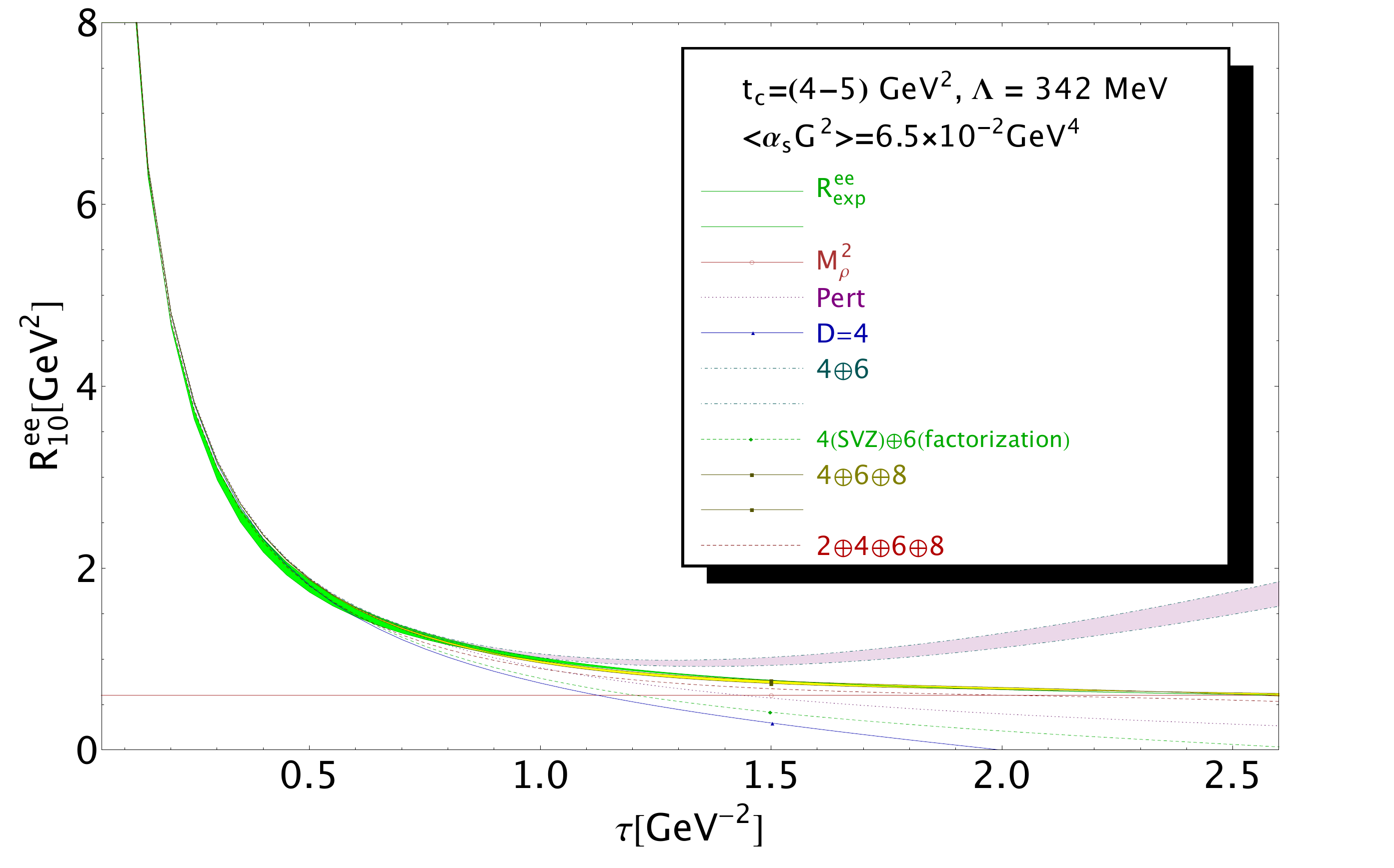}
\caption{\footnotesize  a)  $\tau$-behaviour of the lowest moment ${\cal L}_0$ for different truncation of the OPE; b) The same as in a) but for  the ratio of moments ${\cal R}^{ee}_{10}$.}\label{fig:Ree}.
\end{center}
\vspace*{-0.5cm}
\end{figure} 
\end{center}
\section{QCD versus experiment for ${\cal R}^{ee}_{10}$ and ${\cal L}^{ee}_0$}
Using as input the previous values of the condensates and the value of $\alpha_s$ from PDG, we compare the QCD and experimental sides of  ${\cal R}_{10}$ and ${\cal L}_0$ which we show in Fig.\,\ref{fig:Ree}. The value of the QCD continuum threshold has been fixed by requiring that, for $\tau\leq 0.7$ GeV$^{-2}$ where non-perturbative effecs are expected to be negligible, the PT up to order $\alpha_s^4$ and the experimental sides coincide. This gives:
\beq
t_c\simeq (4\sim 5)~{\rm GeV}^2,
\eeq
which is comparable with the  FESR,\cite{FESR} constraint\,\cite{SNe} 
$
t_c\vert_{fesr}= (5.1\sim 5.6)\,{\rm GeV}^2,
$ 
but much larger than the one of 1.55 GeV$^2$ used in Ref.\,\cite{BOITO,DV} in their estimate of the Duality Violation (DV) from the so-called unpinched moment. Instead, the value used in Ref.\,\cite{BOITO,DV} is obtained when a minimal duality ansatz\,:  ``{$\rho$-meson + $\theta(t-t_c)\times$ QCD continuum"  is used for parametrizing the spectral function. 

\subsection*{\b Ratio of moments ${\cal R}^{ee}_{10}$}
\d For $\tau\leq 1.5$ GeV$^{-2}$, the truncation of the OPE up to $D=6$ gives a good description  of the data where we have a minimum. 

\d The inclusion of the $D=8$ condensate enlarges the region of agreement between the QCD prediction and the experiment until $\tau=2.5$ GeV$^{-2}$ where the $\rho$-meson mass is reached and where, exceptionally, the PT series for the spectral function still make sense. 

\d We complete the analysis by adding the tachyonic gluon mass beyond the SVZ-expansion. We see that it tends to decrease the agreement with the data but the effect is (almost) negligible within our precision. 
\subsection*{\b Lowest moment ${\cal L}^{ee}_{0}$}
We have tried to extract the QCD condensates using ${\cal L}^{ee}_{0}$ but the results are unconclusive. Contrary to the case of 
${\cal R}^{ee}_{10}$, we notice an important sensitivity of the results on the truncation of the PT series. 

\section{QCD condensates from higher $\tau$-like moments in $e^+e^-$}
The analysis has been discussed in details in Ref.\,\cite{SNe2}. 
For instance only $(d_6,d_8)$ contribute to ${\cal R}_0^{ee}$,  $(d_4,d_8,d_{10})$ to ${\cal R}_1^{ee}$, $(d_6,d_{10},d_{12})$ to ${\cal R}_2^{ee}$...Using as input the value of $\alpha_s$ in the PT contributions, we notice that a two-parameter fit of ${\cal R}_0^{ee}$
does not provide a stable result versus the change of $\tau$-like mass. From ${\cal R}_1^{ee}$, one can extract $d_{10}$ by using as input $d_4$ and $d_6$. We proceed in a similar way for higher moments and obtain the results quoted in Table\,\ref{tab:cond}.
The analysis of the $\tau$-like mass $M_0$-stability is illustrated in Figs.\,\ref{fig:d8} and \,\ref{fig:d14}. 

\begin{figure}[hbt]
\begin{center}
\hspace*{-4cm}{\bf a)}\hspace*{8cm} {\bf b)}\\
\includegraphics[width=7.5cm]{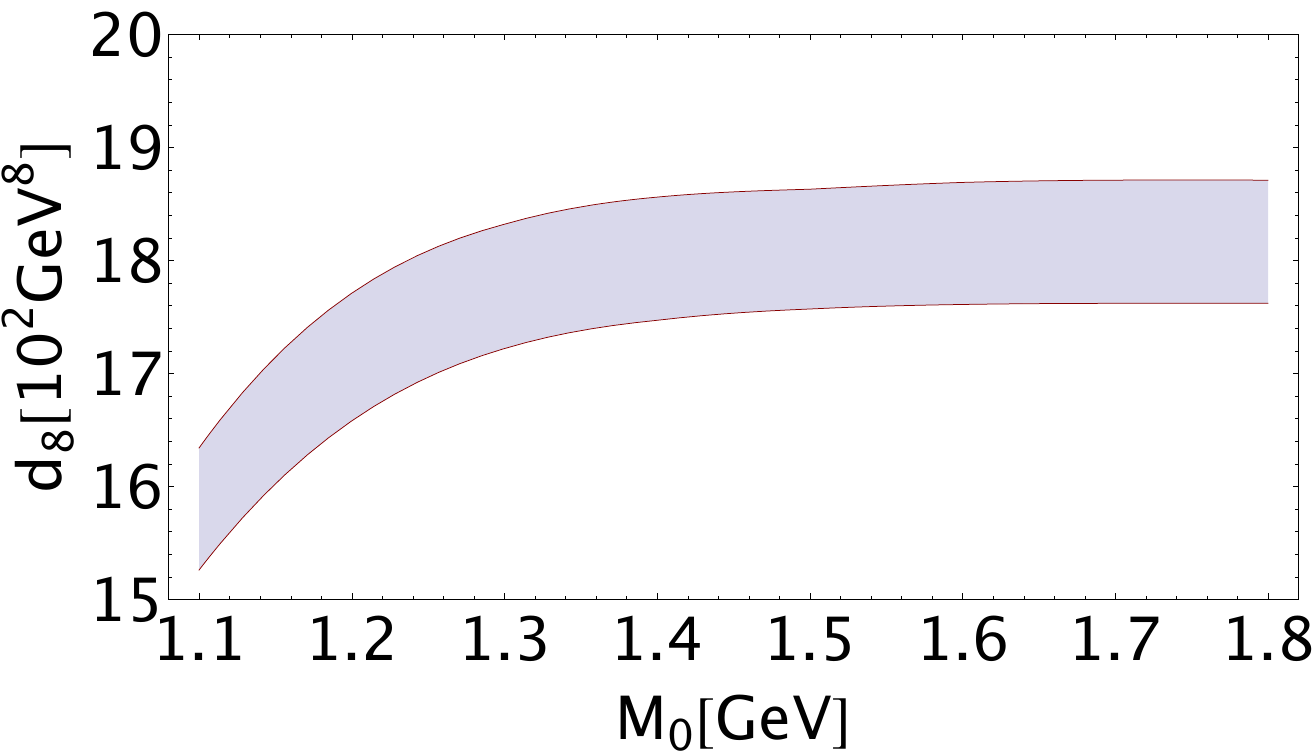}
\includegraphics[width=7.5cm]{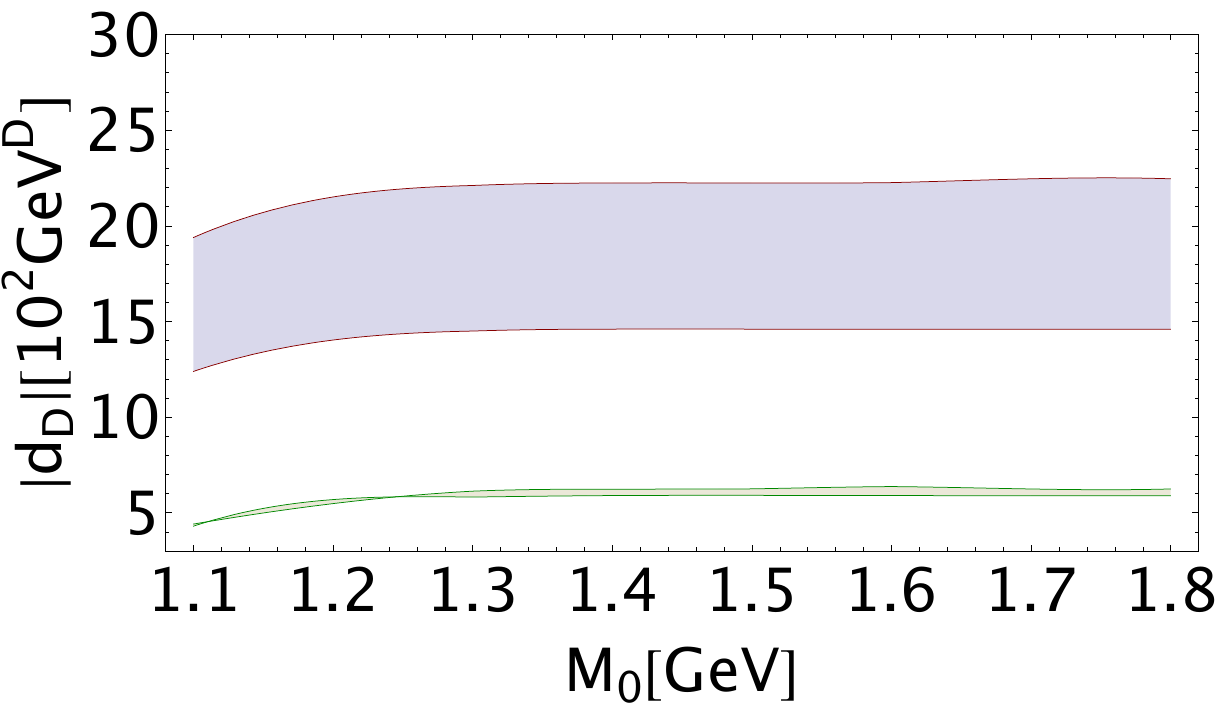}
\caption{\footnotesize {\bf a)} Value of the condensate $d_8$ from ${\cal R}_1^{ee}$ versus $M_0$;
{\bf b)} Values of the condensates $d_{10}$ and $d_{12}$ from ${\cal R}_1^{ee}$ and 
${\cal R}_2^{ee}$  versus $M_0$}\label{fig:d8}
\end{center}
\vspace*{-0.5cm}
\end{figure}  

\begin{figure}[hbt]
\begin{center}
\hspace*{-4cm}{\bf a)}\hspace*{8cm} {\bf b)}\\
\includegraphics[width=7.5cm]{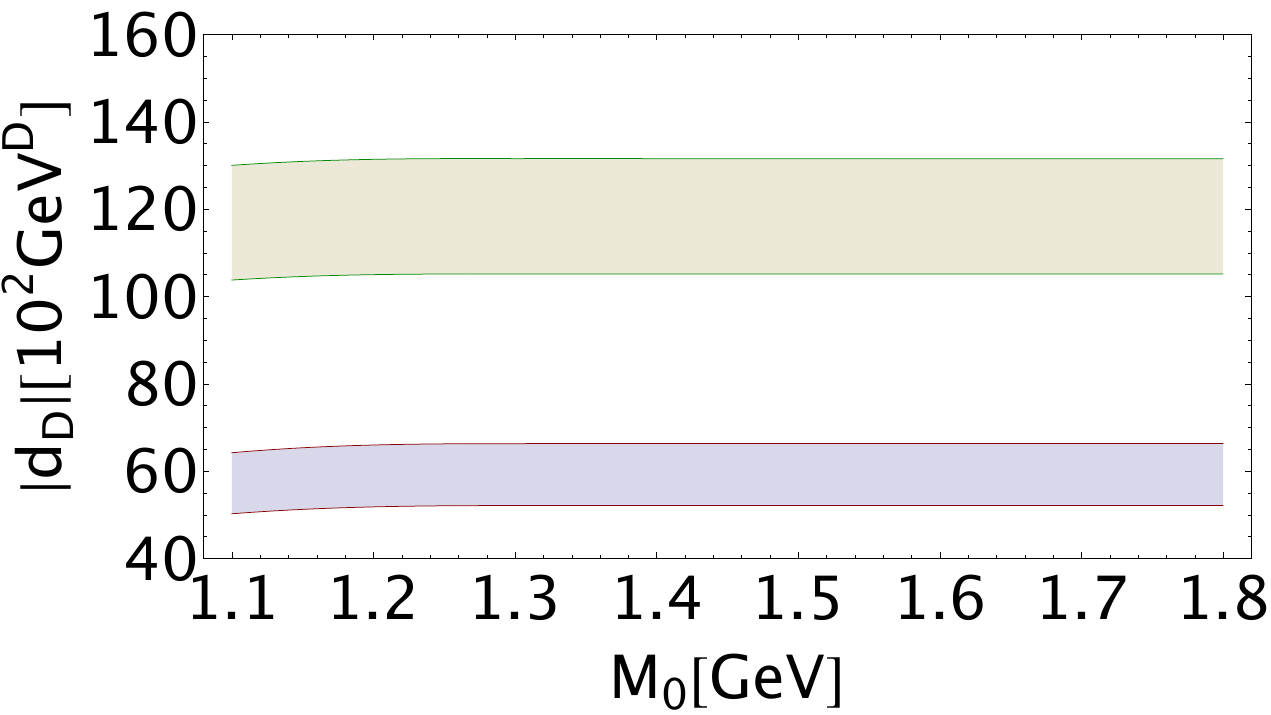}
\includegraphics[width=7.5cm]{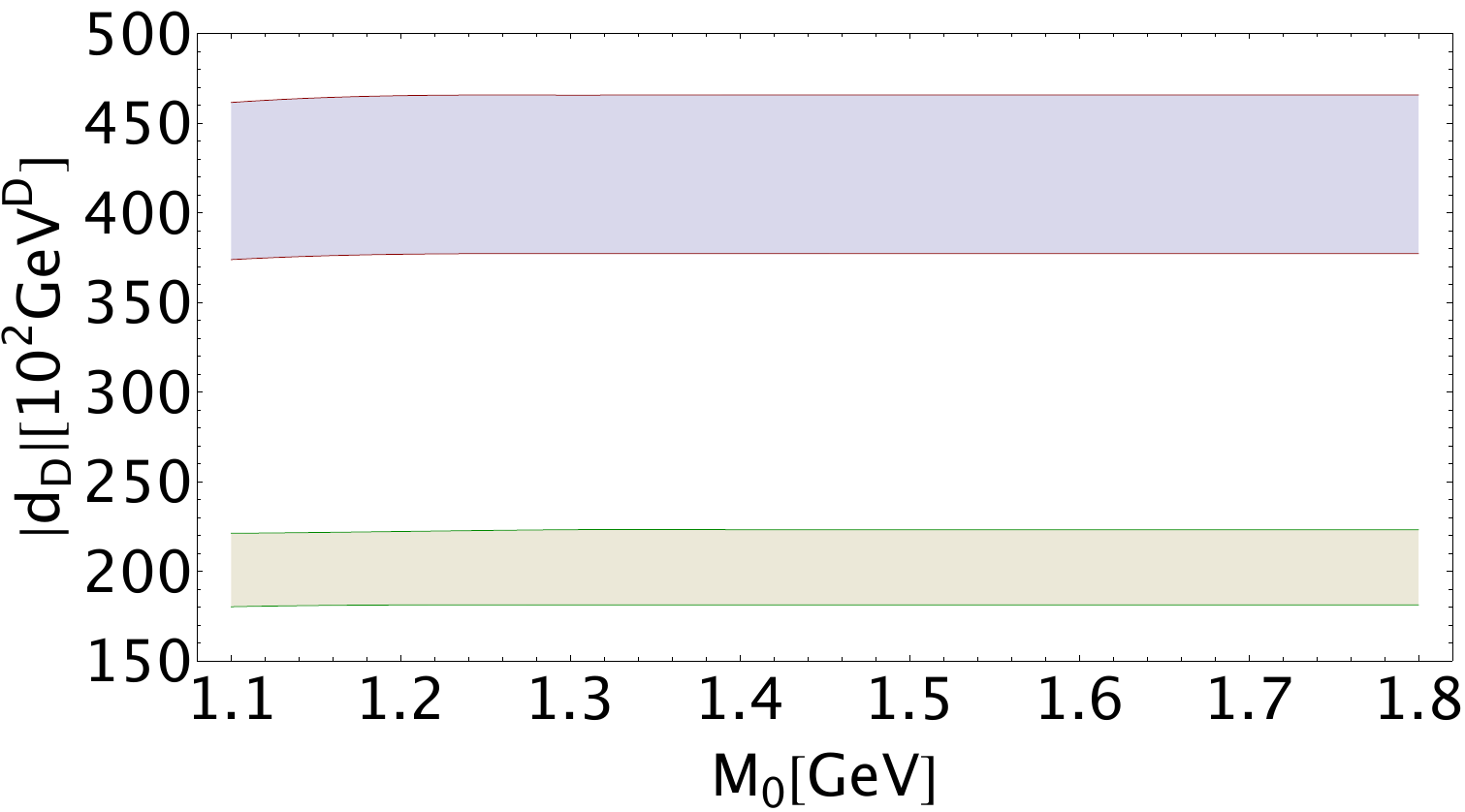}
\caption{\footnotesize {\bf a)} Value of the condensate $d_{14}$ and $d_{16}$ from ${\cal R}_3^{ee}$
and ${\cal R}_4^{ee}$  versus $M_0$;
{\bf b)} Value of the condensates $d_{18}$ and $d_{20}$  from ${\cal R}_5^{ee}$ and 
${\cal R}_6^{ee}$  versus $M_0$}\label{fig:d14}
\end{center}
\end{figure}  

   {\scriptsize
\begin{table}[hbt]
\setlength{\tabcolsep}{0.15pc}
  \begin{center}
    {\footnotesize  \begin{tabular}{lllll ll ll}

&\\
\hline
$\oliva \la\alpha_s G^2\ra$&$\oliva -d_6$&$\oliva d_8$&-$\oliva d_{10}$&-$\oliva d_{12}$&$\oliva d_{14}$& $\oliva -d_{16}$&$\oliva -d_{18}$&$\oliva d_{20}$\\
 \hline 
$7.8\pm 3.5$&$23.6\pm 3.7$&$18.2\pm 0.6$&$6.1\pm 0.2$&$18.4\pm 3.8$&$59.2\pm 7.1$&$118.3\pm 13.2$&$ 202.0\pm 20.7$&$421.3\pm 44.3$\\
   \hline
\end{tabular}}
 \caption{\footnotesize Values of the QCD condensates of dimension $D$ in units of $10^{-2}$ GeV$^{D}$ from  the present analysis.}\label{tab:cond} 
 \end{center}
\end{table}
} 
 \section{Re-estimate of $\la\alpha_s G^2\ra$ and $d_6$ from the ratio ${\cal R}^{ee}_{10}$}
 \d Using the previous values of the $d=8$ to $d=20$ condensates into the ratio ${\cal R}_{10}$ of LSR, 
we re-estimate $d_6$ and $d_4$. To strongly constrain the parameters,
we continue to use a one-parameter fit by fixing the value of $\la \alpha_s G^2\ra$ as in Eq.\,\ref{eq:asg2} and we re-extract $d_6$ by itruncating the OPE in the QCD expression of ${\cal R}_{10}$ up to $D=20$ dimension. The analysis is shown  in Fig.\,\ref{fig:d6} where one can notice that the $\tau$-stabilities are obtained at small values of $\tau$.  These vaalues are quoted in Table\,\ref{tab:cond}. One can notice that the value of $\la\alpha_s G^2\ra$ is less accurate than the one quoted in Eq.\,\ref{eq:asg2}. These values supersed the ones obtained in Ref.\,\cite{SNe} at larger values of $\tau$ and using only the OPE truncated at $D=8$.
\begin{figure}[hbt]
\begin{center}
\hspace*{-4cm}{\bf a)}\hspace*{8cm} {\bf b)}\\
\includegraphics[width=7.5cm]{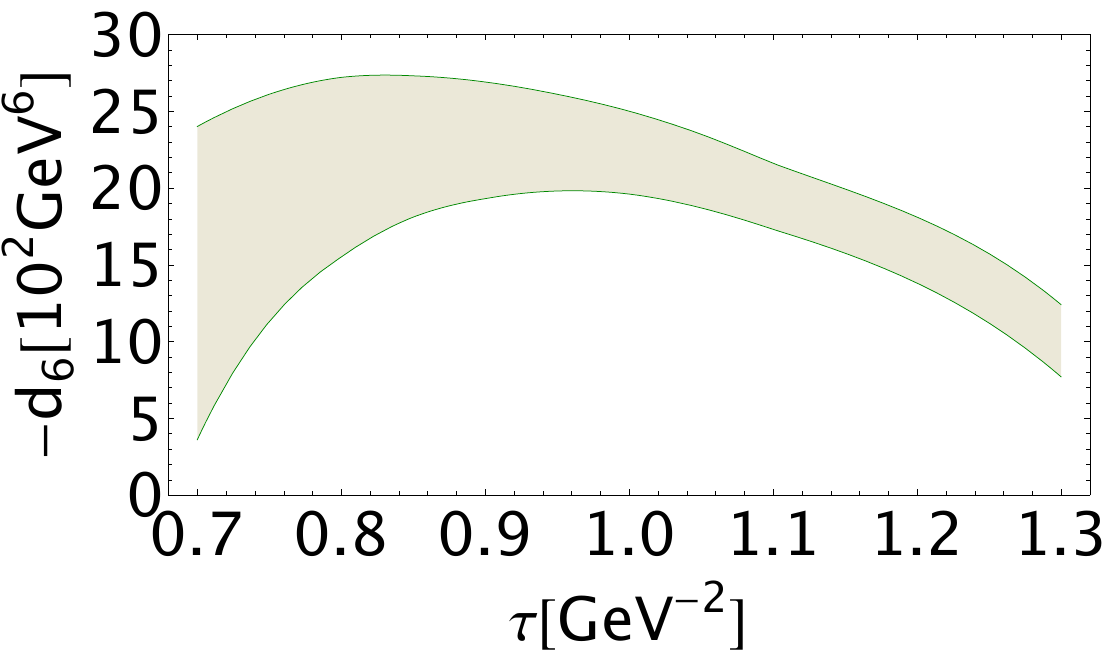}
\includegraphics[width=7.5cm]{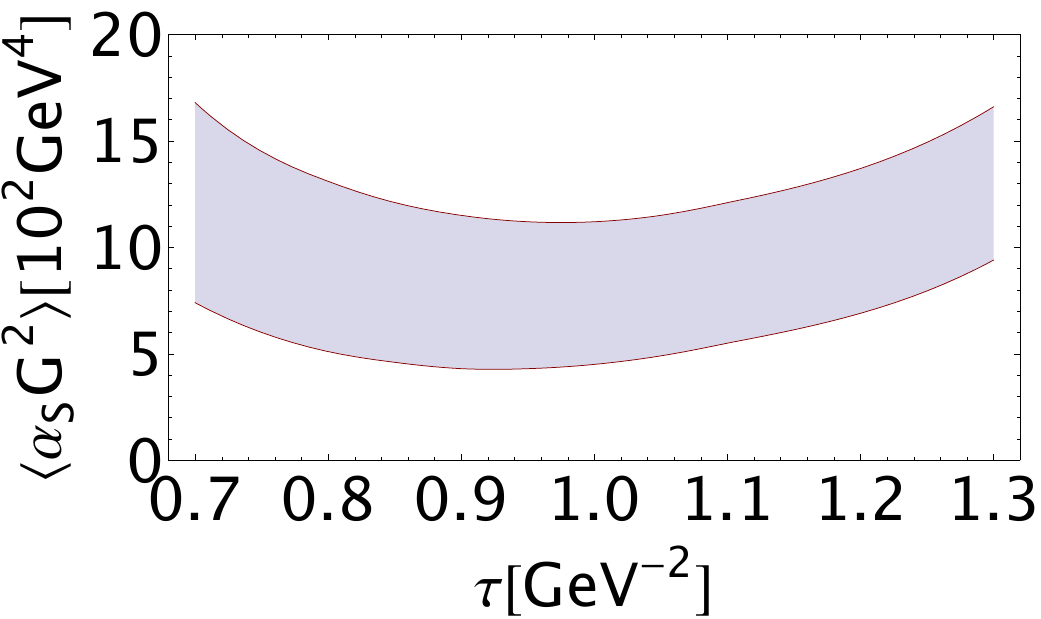}
\caption{\footnotesize {\bf a)} Value of the condensate $d_6$  from ${\cal R}_{10}^{ee}$
versus $M_0$;
{\bf b)} Value of the condensate $d_{4}$   from ${\cal R}_{10}^{ee}$   versus $M_0$}\label{fig:d6}
\end{center}
\end{figure}  

\d One can notice that the ratio\,:
\beq
r_{46} \equiv \frac{\rho\la\bar\psi\psi\ra^2}{\la\alpha_s G^2\ra}\approx 0.9\times 10^{-2}~{\rm GeV^2},
\eeq
is almost constant as obtained here  and from different approaches\,\cite{LNT,FESR,SOLA,BORDES,CAUSSE} independently on the absolute size of the condensates and on the methods used to extract them.

\begin{figure}[H]
\begin{center}
\includegraphics[width=7.5cm]{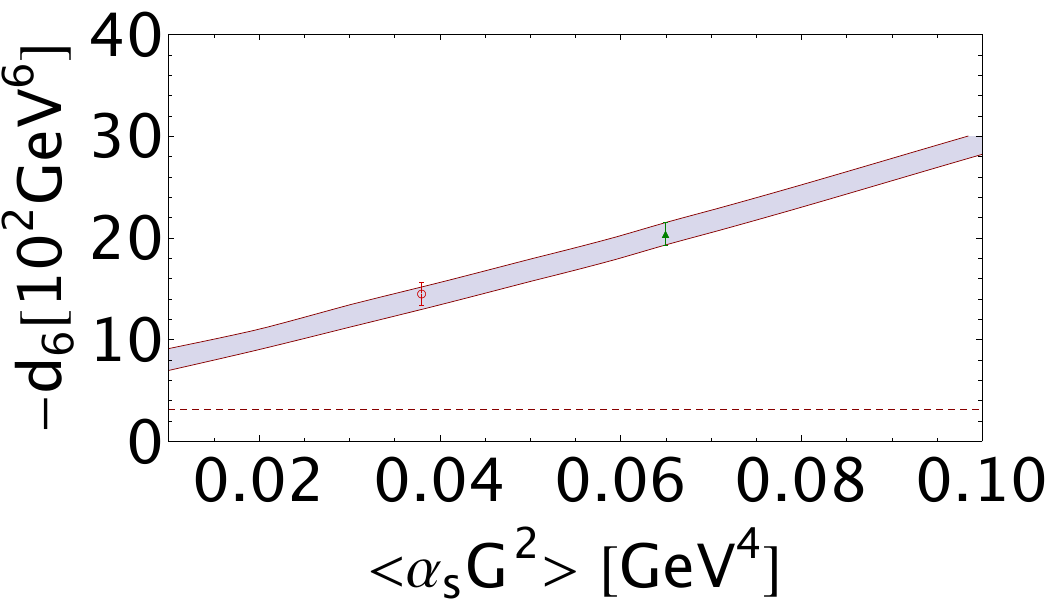}
\caption{\footnotesize  Correlated values of $d_6$ versus $\la\alpha_s G^2\ra.$ The dashed horizontal line is the value of $d_6$ estimated from factorization of the four-quark condensate. The red (resp. oliva) point corresponds to the value of $\la\alpha_s G^2\ra$ given by SVZ and by Eq.\,\ref{eq:asg2}. } \label{fig:d68-g2}.
\end{center}
\vspace*{-0.5cm}
\end{figure} 
  \subsection*{\hspace*{0.5cm} \d Correlated values of  $\la\alpha_s G^2\ra$ and $d_6$ }
 We show in Fig.\,\ref{fig:d68-g2} the behaviour of $d_6$ versus $\la\alpha_s G^2\ra$. One can notice that the value of the four-quark condensate estimated from factorization is inconsistent with the SVZ value of the gluon condensate.
 \vspace*{-1cm}
   {\scriptsize
   \begin{center}
\begin{table}[hbt]
\setlength{\tabcolsep}{0.5pc}
  \begin{center}
    {\footnotesize
  \begin{tabular}{lllll ll}

&\\
\hline
\oliva$\la\alpha_s G^2\ra$ &\oliva$-d_6$ &\oliva$d_8$&\oliva$-d_{10}$&\oliva$-d_{12}$  &\oliva Refs.\\
 \hline 
$ 7.8\pm 3.5$&$23.6\pm 3.7$&$18.2\pm 0.6$&$6.1\pm 0.2$&$18.4\pm 3.8$& This work \\
$0.67\pm 0.89$&$15.2\pm 2.2$&$22.3\pm 2.5$&&& {\rm ALEPH\,\cite{ALEPH1}}\,\\
$5.34\pm 3.64$&$14.2\pm 3.5$&$21.3\pm 2.5$&&& {\rm OPAL\,\cite{OPAL}}\,\\
 $3.5^{+2.2}_{-3.8}$&$19.7^{+11.8}_{-7.9}$&$23.7^{+11.8}_{-15.8}$&$11.8\pm 19.7$&$7.9\pm 19.7$&(${  d_{14,16}=0}$){  Pich-Rodriguez\,\cite{PICH1}}\\

   \hline
\end{tabular}}

 \caption{Values of the QCD condensates of dimension $D$ in units of $10^{-2}$ GeV$^{D}$  in the Vector channel from some other $\tau$-moments at Fixed Order (FO) of the PT series.}\label{tab:other} 
 \end{center}
\end{table}
\end{center}
} 
\section{$\alpha_s$ from the lowest  BNP $\tau$-like moment ${\cal R}_0^{ee}$ in $e^+e^-$}
\d Once fixed the values of the condensates, we attempt to extract $\alpha_s$ from the 
lowest ${\cal R}_0^{ee}$ BNP moment\,\cite{BNP} which is less affected by the lowest condensate contribution
as the one of  $\la\alpha_s G^2\ra$  vanishes to leading order in $\alpha_s$. With the set of condensates in Table\,\ref{tab:cond}, one shows,  in Fig.\,\ref{fig:as}, the behaviour of $\alpha_s$ versus the hypothetical $\tau$-mass $M_0$\,\cite{SNe2}. 

\begin{figure}[hbt]
\begin{center}
\includegraphics[width=12cm]{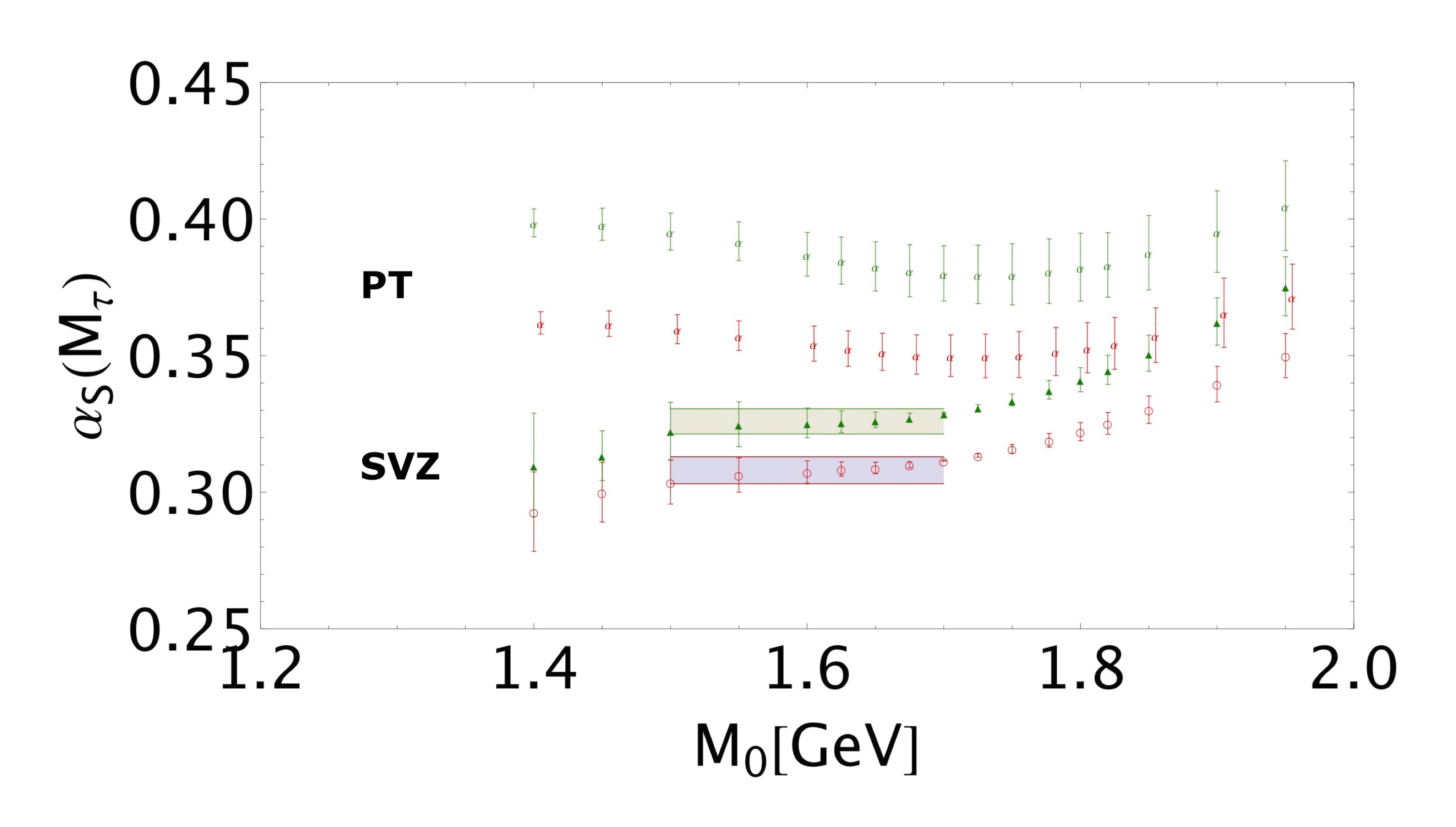}
\caption{\footnotesize {\bf a)} Value of $\alpha_s (M_\tau)$ from the BNP lowest moment
for different values of $M_0$. The red (green) points correspond to (FO) [resp. (CI)] PT series.
The lower curves correspond to the SVZ expansion. The upper ones to the case of zero value of the QCD condensates.
}\label{fig:as}
\end{center}
\vspace*{-0.5cm}
\end{figure}  

\d One can notice that the stable value is obtained in the region $M_0=(1.5-1.7)$\,GeV but not at the physical $M_\tau$ where the value is overestimated. Including the higher order (HO) corrections via the estimate of the $\alpha_s^5$ contribution, one obtains\,\cite{SNe2}:
\bea
\alpha_s(M_\tau)\vert_{e^+e^-}^{SVZ}  &=& 0.3081(49)_{fit}(71)_{\alpha_s^5} \lrar2~\alpha_s(M_Z)\vert_{e^+e^-}^{SVZ}  = 0.1170(6)(3)_{evol} ~   {\rm (FO)}\nnb\\
&=& 0. 3260(47)_{fit}(62)_{\alpha_s^5} \lrar2~\alpha_s(M_Z)\vert_{e^+e^-}^{SVZ}  = 0.1192(6)(3)_{evol} ~{\rm (CI)},
\label{eq:as-foci}
\eea
for Fixed Order (FO) and Cointour Improved (CI) perturbative series, 
while the sum of the non-perturbative corrections at the $\tau$-mass is :  $\delta^{(NP)}= (3.8\pm 0.8) \times 10^{-2}$ which is  negligible. The value in Eq.\,\ref{eq:as-foci} is slightly lower than the one in Ref.\,\cite{SNe} obtained with some other set of condensates. 

\d One can also notice that a QCD Model without QCD condensates give too high value of $\alpha_s$:
\beq
\alpha_s(M_\tau)\vert_{e^+e^-}^{PT}   =  0.3579(115)_{fit}(34)_{\alpha_s^5} ~~~\lrar2~~~\alpha_s(M_Z)\vert_{e^+e^-}^{PT}  = 0.1228(12)(3)_{evol} ~~~~~~   {\rm (FO)}, 
\eeq
which is not favoured by the world average\,\cite{PDG}:
 \beq
\la\alpha_s(M_Z)\ra= 0.1178(5).
\label{eq:pdg}
\eeq
\d Some different estimates from the Vector component of $\tau$-decay using 
higher moments are shown in Table\,\ref{tab:tau}. We also show our prediction at $M_\tau$ for the V component of $\tau$-decay using $d_6$ and $d_8$ in Table\,\ref{tab:cond}. We have used the updated ALEPH data \,\cite{ALEPH}:
\beq
{\cal R}^{exp}_{\tau,V}= 1.782\pm 0.009.
\eeq

\d For a better comparison with our result where the condensates have been estimated within (FO), we quote the value of $\delta^{(NP)}$ corresponding to (FO) at $M_\tau$.


  {\footnotesize
     \begin{center}
\begin{table*}[hbt]
\setlength{\tabcolsep}{0.2pc}
\newlength{\digitwidth} \settowidth{\digitwidth}{\rm 0}
\catcode`?=\active \def?{\kern\digitwidth}
\label{tab:effluents}
\footnotesize{
\begin{tabular*}{\textwidth}{@{}l@{\extracolsep{\fill}}cccccc c}
\hline
              & \multicolumn{2}{c}{\oliva THIS WORK} 
                  &\oliva ALEPH\,\cite{ALEPH1}&\oliva OPAL\,\cite{OPAL} &\oliva PR\,\cite{PICH1}&\oliva ALEPH\,\cite{ALEPH}\\
\cline{2-3} 
                 & \multicolumn{1}{c}{$e^+e^-$} 
                 & \multicolumn{1}{c}{$\tau$-decay} \\
                   \hline

FO&0.3081(86)&0.3128(79)& 0.3200(220)&0.3230(160)&0.3200(150)&--&\\
CI & 0.3260(78)&0.3291(70)&0.3400(230)&0.3470(220)&0.3370(200)&0.3460(110) \\
$\delta^{(NP)}\,\times 10^{2}$&$3.8\pm 0.8$&$3.7\pm1.0$&$2.0\pm 0.3$&$3.6\pm 0.4$&$1.7\pm 0.3$&$2.0\pm 0.3$&\\
\hline
\vspace*{-0.5cm}
\end{tabular*}
}
 \caption {\footnotesize  $\alpha_s(M_\tau)$ within the SVZ-expansion from the vector (V) component of $\tau$-decay data using some high-$\tau$-moments  and our results from $e^+e^-$. }\label{tab:tau}
\end{table*}
 \end{center}
}

\section{Beyond the SVZ-expansion\label{sec:beyond} }
 We shall not include some contributions beyond the SVZ expansion namely\,:
 
\d The $D=2$ contribution due to an eventual tachyonic gluon mass which has been introduced in Ref.\,\cite{CNZ} to parametrize phenomenologically the contribution of UV renormalon in the PT series\,\footnote{For an analysis using a large $\beta$-approximation see  e.g.\,\cite{CVETIC}.}. However, it is
expected that contribution is dual to the estimate non-calculated Higher Order terms,\cite{SNZ} which we assume to be mimiced  by our estimate of $\alpha_s^5$.
  
\d Instanton which are expected to negligible\,\cite{SNe}. 

\d Duality violation (DV)\,\cite{BOITO,DV} shown in\,\cite{PICH1} to be negligible. We reach the same conclusion from the analysis in Ref.\,\cite{SNe,SNe2}. 

\section{The axial-vector (A) component of $\tau$-decay\label{sec:axial} }
We use the same strategy as in the case of $e^+e^-$ to fit the data of the axial-vector spectral function given in Fig.\,\ref{fig:aleph}. To  check our fit, we compare the experimental value of 
the lowest $\tau$-decay moment ${\cal R}_{0,A}(\tau)$ at the $\tau$-mass with the one measured by ALEPH\,\cite{ALEPH} :
\beq
{\cal R}_{0,A}\vert_{\rm our\, fit}= 1.698(14),~~~~~ {\cal R}_{0,A}\vert_{Aleph} = 1.694(10),
\eeq
\subsection*{\b Determination of the QCD condensates from the ratio $R^A_{10}$ of LSR}

\d We use the ratio of LSR moment to extract $(d_{6,A},d_{8,A})$ from a two-parameter fit giving $\la\alpha_s G^2\ra$ as input. The $\tau$-behaviour of the result is given in Fig.\,\ref{fig:r10}.
\begin{figure}[hbt]
\begin{center}
\hspace*{0.cm}{\bf a)} \hspace*{6cm}{\bf b)}\\
\includegraphics[width=7.1cm]{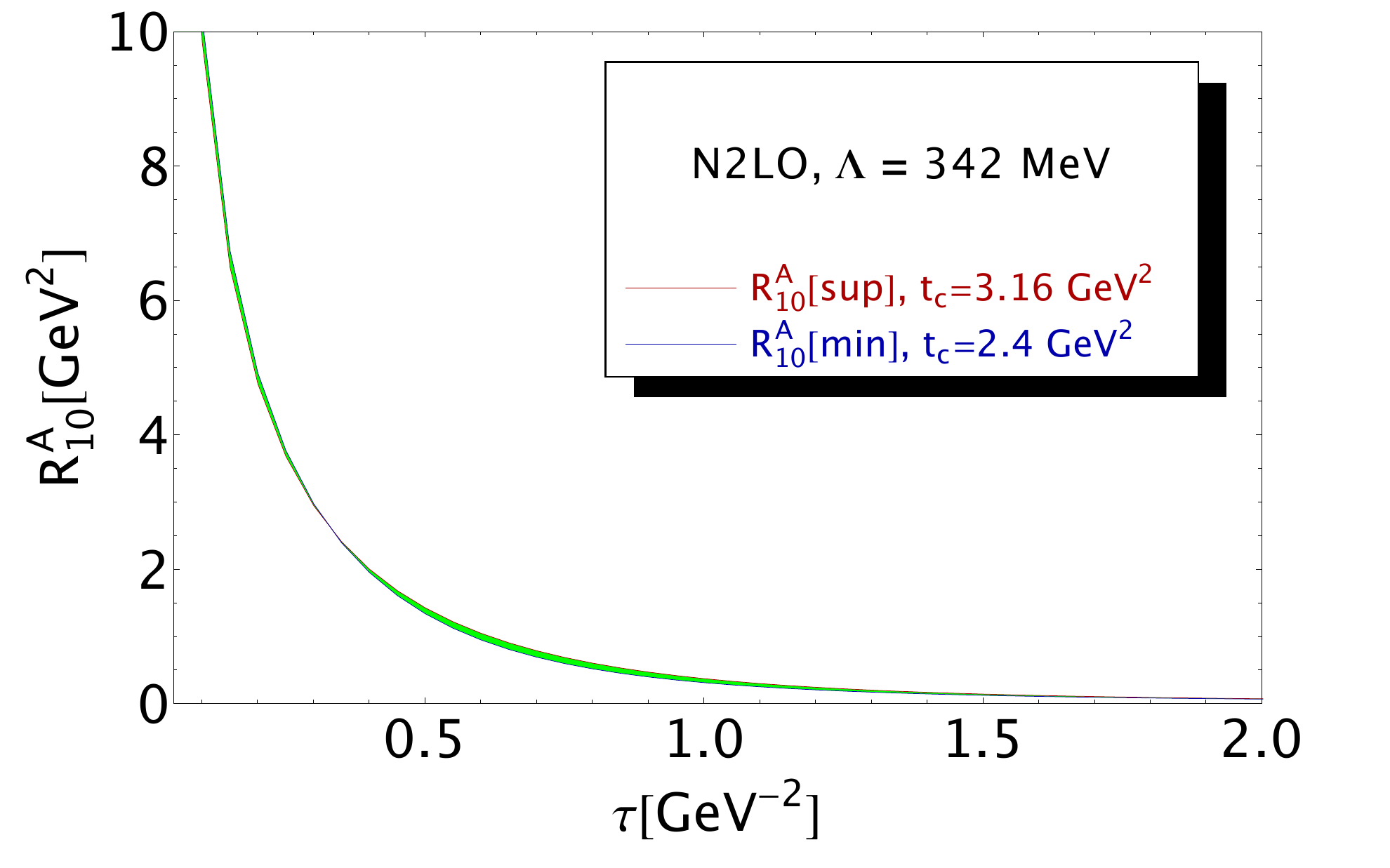}
\includegraphics[width=8.cm]{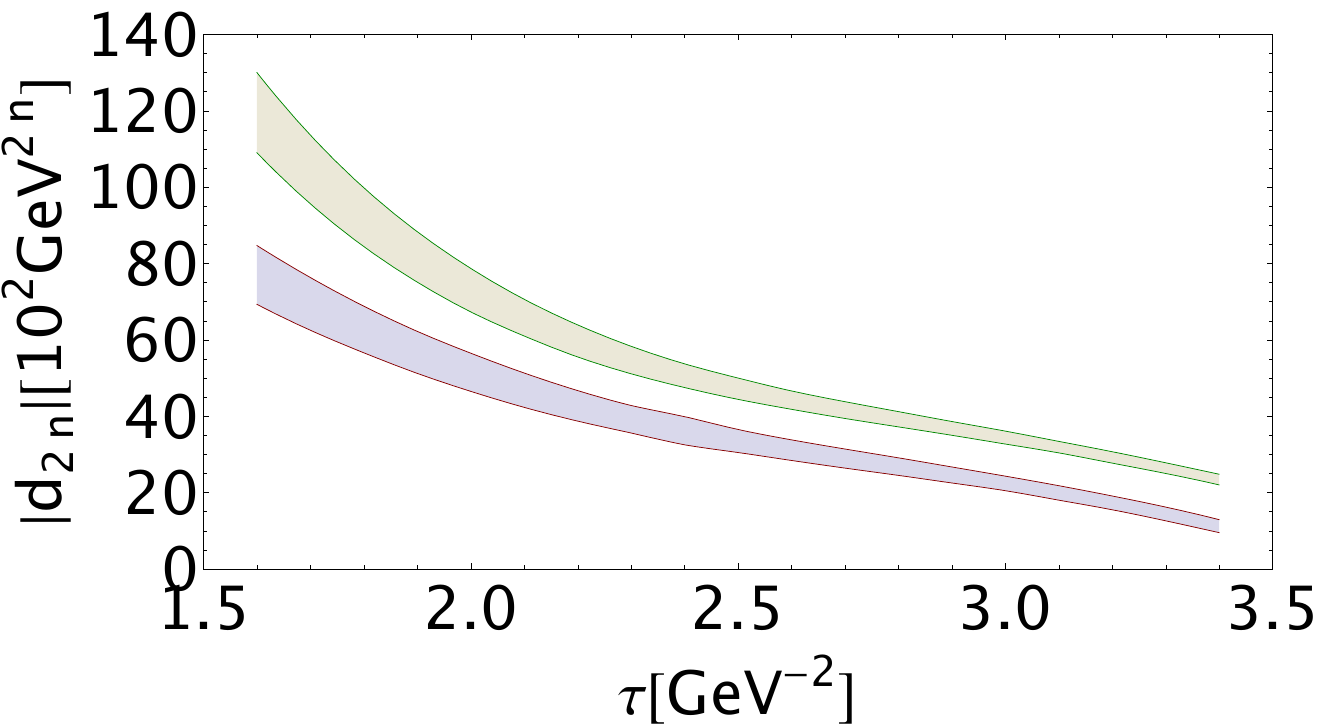}
\caption{\footnotesize {\bf a)}: $R^A_{10}$ versus the LSR variable $\tau$;  {\bf b)}:  $d_{6,A}$ and $d_{8,A}$ from the $ {\cal R}^A_{10}$} \label{fig:r10}
\end{center}
\vspace*{-0.5cm}
\end{figure} 

In this way, we obtain at the inflexion point $\tau\simeq (2.5\pm 0.1)$ GeV$^{-2}$:
\beq
d_{6,A} =  (33.5\pm 3.0\pm 2.7)\times 10^{-2}\,{\rm GeV^6} ~ ~~~~~~~~~d_{8,A} = -(47.2\pm 2.8\pm 3.2)\times 10^{-2}\,{\rm GeV^8}
\label{eq:d68-ratio}
\eeq
where the errors come respectively from the fitting procedure and the localization of the inflexion point. 

\d We redo the fit by fixing now $(d_{6,A},d_{8,A})$ and extract $\la\alpha_s G^2\ra$. The analysis is shown in Fig.\\ref{fig;g2},
\begin{figure}[hbt]
\begin{center}
\includegraphics[width=9cm]{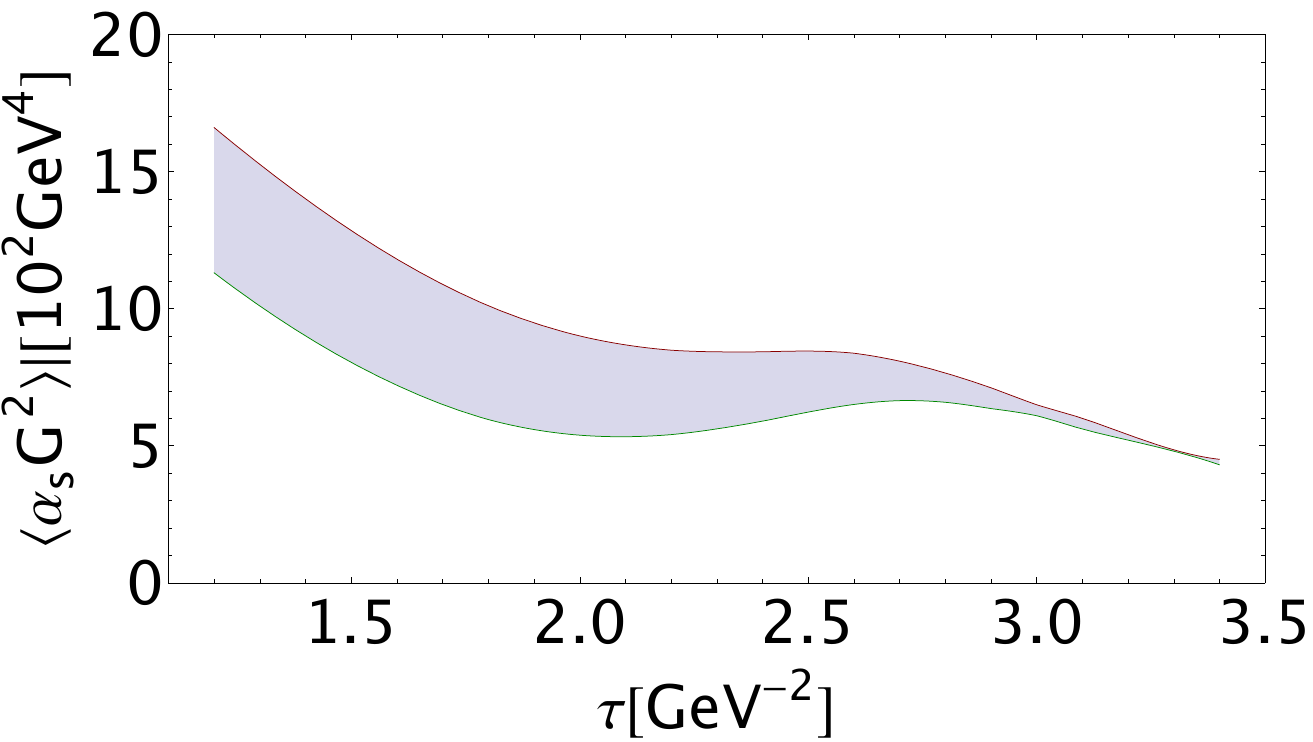}
\caption{\footnotesize  $\la \alpha_s G^2\ra$ versus the LSR variable $\tau$. } \label{fig:g2}
\end{center}
\vspace*{-0.5cm}
\end{figure} 
from which, we deduce:
\beq
\la \alpha_s G^2\ra =(6.9\pm 1.5)\times 10^{-2}\,{\rm GeV}^4,
\eeq
in good agreement with the one in Eq.\,\ref{eq:asg2} from the heavy quark systems and some other sum rules used previously as input though less accurate.
\subsection*{\b The QCD condensates from higher ${\cal R}_{n,A}$ moments}
The analysis is done in details in Ref.\,\cite{SNtau24}. We show in fig.\,\ref{fig:d6-8} the $s_0=M_0^2$ hypothetical $\tau$-mass squared behaviour of the $d_{6,A}$ to $d_{10,A}$ where a nice $s_0$ stability is observed, at which, we deduce the optimal results.
\begin{figure}[hbt]
\begin{center}
\hspace*{-4cm} {\bf a)} \hspace*{8cm}{\bf b)}\\
\includegraphics[width=7.5cm]{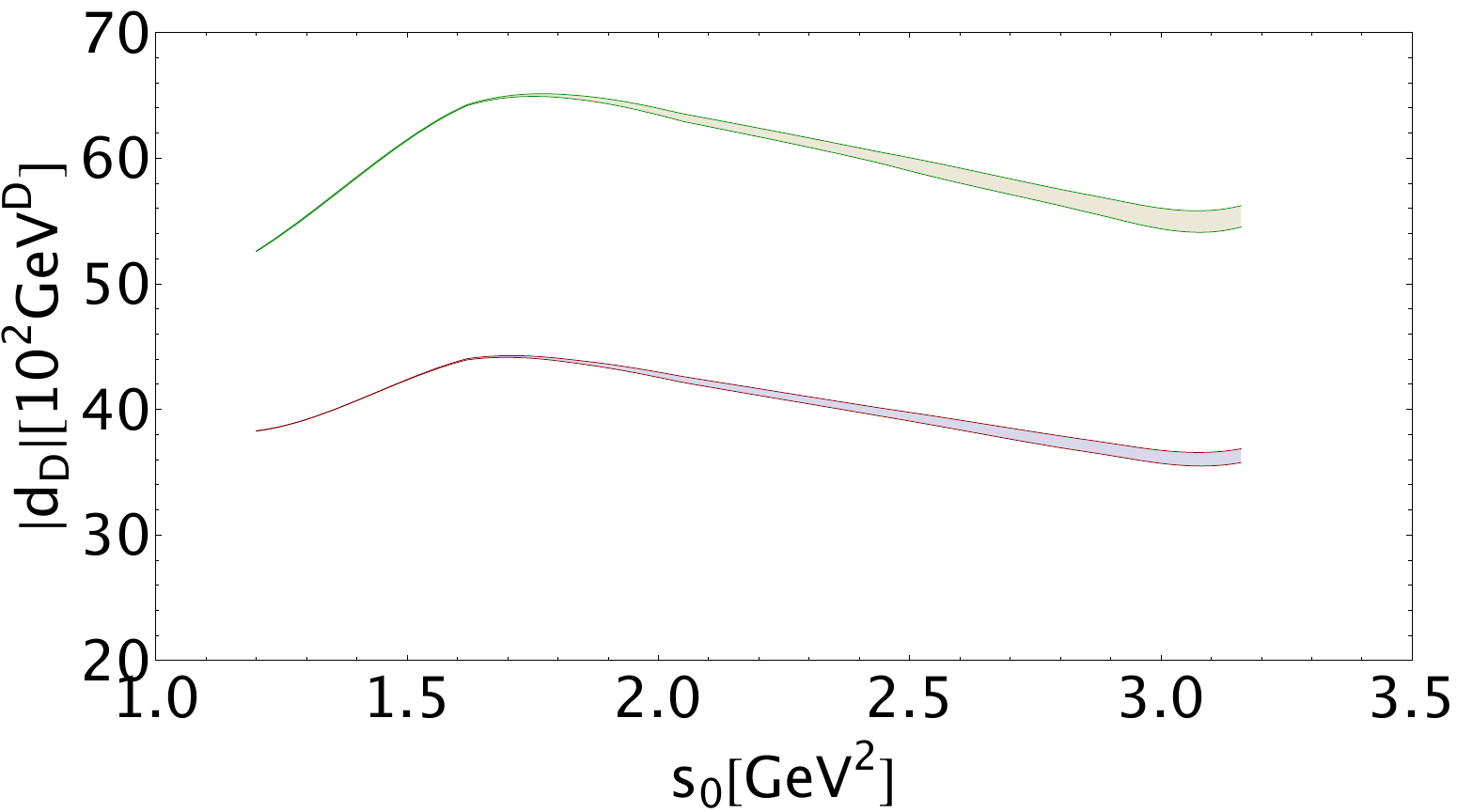}
\includegraphics[width=7.5cm]{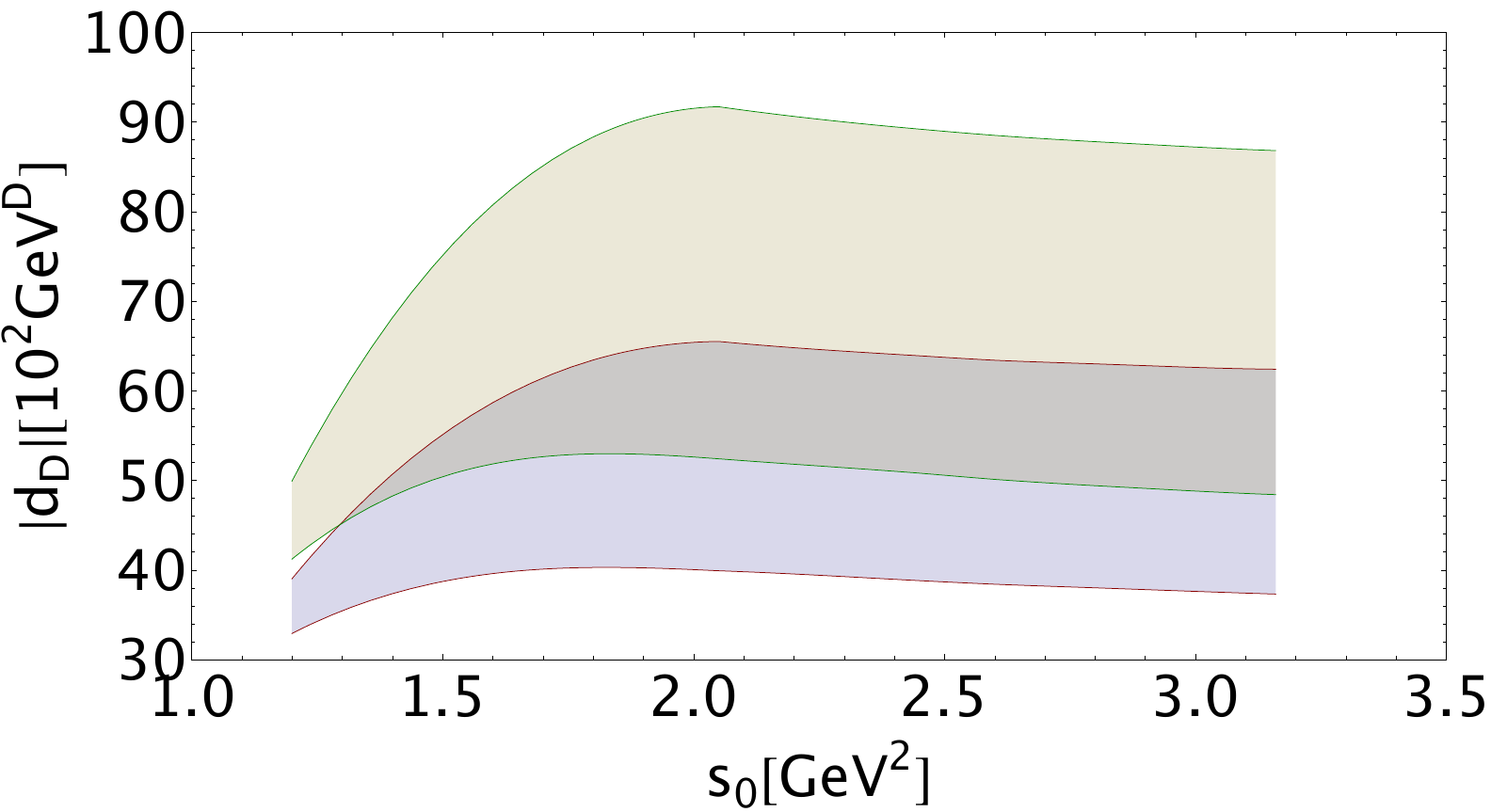}
\caption{\footnotesize  {\bf a)}: $|d_{6,A}|$ (lowest curve) and $|d_{8,A}|$ (highest curve)  versus  $M^2_0$; {\bf b)}: similar to  {\bf a)} but for $|d_{8,A}|$ and $|d_{10,A}|$. } \label{fig:d6-8}
\end{center}
\vspace*{-0.5cm}
\end{figure} 
The final combined results from the ratio $R^A_{10}$ of LSR and  ${\cal R}_{n,A}$ moments are given in Table\,\ref{tab:other}  and compared with some other determinations.
   {
\begin{table}[H]
\setlength{\tabcolsep}{0.1pc}
  \begin{center}
    {\footnotesize
  \begin{tabular}{lllll ll ll}

&\\
\hline
$\oliva d_{6,A}$&$\oliva -d_{8,A}$&$\oliva d_{10,A}$&$\oliva -d_{12,A}$&$\oliva d_{14,A}$& $\oliva -d_{16,A}$&$\oliva d_{18,A}$&$\oliva- d_{20,A}$&Refs.\\
 \hline 
$34.4\pm 1.7$&$51.5\pm 2.1$&$53.9\pm 0.7$&$63.3\pm1.8$&$77.0\pm 6.0$&$93.1\pm 4.0$&$104.3\pm 7.7$&$ 119.7\pm 8.2$& This work\\
$43.4\pm 13.8$&$59.2\pm 19.7$&$63.2\pm 33.6$&$43.4\pm31.6$&&&&&\cite{PICH1}\\
 $19.7\pm 1.0$&$27\pm 1.2$&&&&&&&\cite{ALEPH} \\
 $9.6\pm 3.3$&$9.0\pm 5.0$&&&&&&&\cite{OPAL} \\
   \hline
\end{tabular}}
 \caption{ Values of the QCD condensates of dimension $D$ in units of $10^{-2}$ GeV$^{D}$ from  this work and some other estimates. }\label{tab:cond-A} 
 \end{center}
\end{table}
} 
\d  We notice,  by comparing the results from $e^+e^-$ and the Axial-vector channel, that the relation :
\beq
d_{6,A}\simeq  -(11/7)\,d_{6,V},
\eeq
 is quite well satisfied within the errors.   This result also suggests a violation of the four-quark condensate vacuum saturation (see Eq.\,\ref{eq:d8})  similar to the one found from $e^+e^-\to$ Hadrons data\,\cite{SNe,SNe2}\,:
 \beq
 \rho\alpha_s\la\bar \psi\psi\ra^2 = (6.38\pm 0.30)\times 10^{-4}\,{\rm GeV}^6~~~~ \lrar2 ~~~~ \rho\simeq  (6.38\pm 0.30).
 \label{eq:d8}
 \eeq

\d Like in the case of vector channel, we do not observe any exponential growth of the size of the condensates which indicates that one also expects a small effect of the duality violation (DV) in the axial-vector channel. However, in this channel, it is remarkable that the condesate contributions have alternate signs. 
\subsection*{\b Determination of $\alpha_s(M_\tau)$ from the Axial-vector channel }
We show in Fig.\,\ref{fig:as-A}, the behaviour of $\alpha_s(M_\tau)$ versus an hypothetical $\tau$ mass squared $M_0^2\equiv s_0$. One can notice an inflexion point in the region $2.5^{+0.10}_{-0.15}$ GeV$^2$ at which we extract the optimal result. The conservative result
from $s_0=2.1$ GeV$^2$ to $M_\tau^2$ (see Fig.\,\ref{fig:as-A}) is\,:
\begin{figure}[hbt]
\begin{center}
\includegraphics[width=10cm]{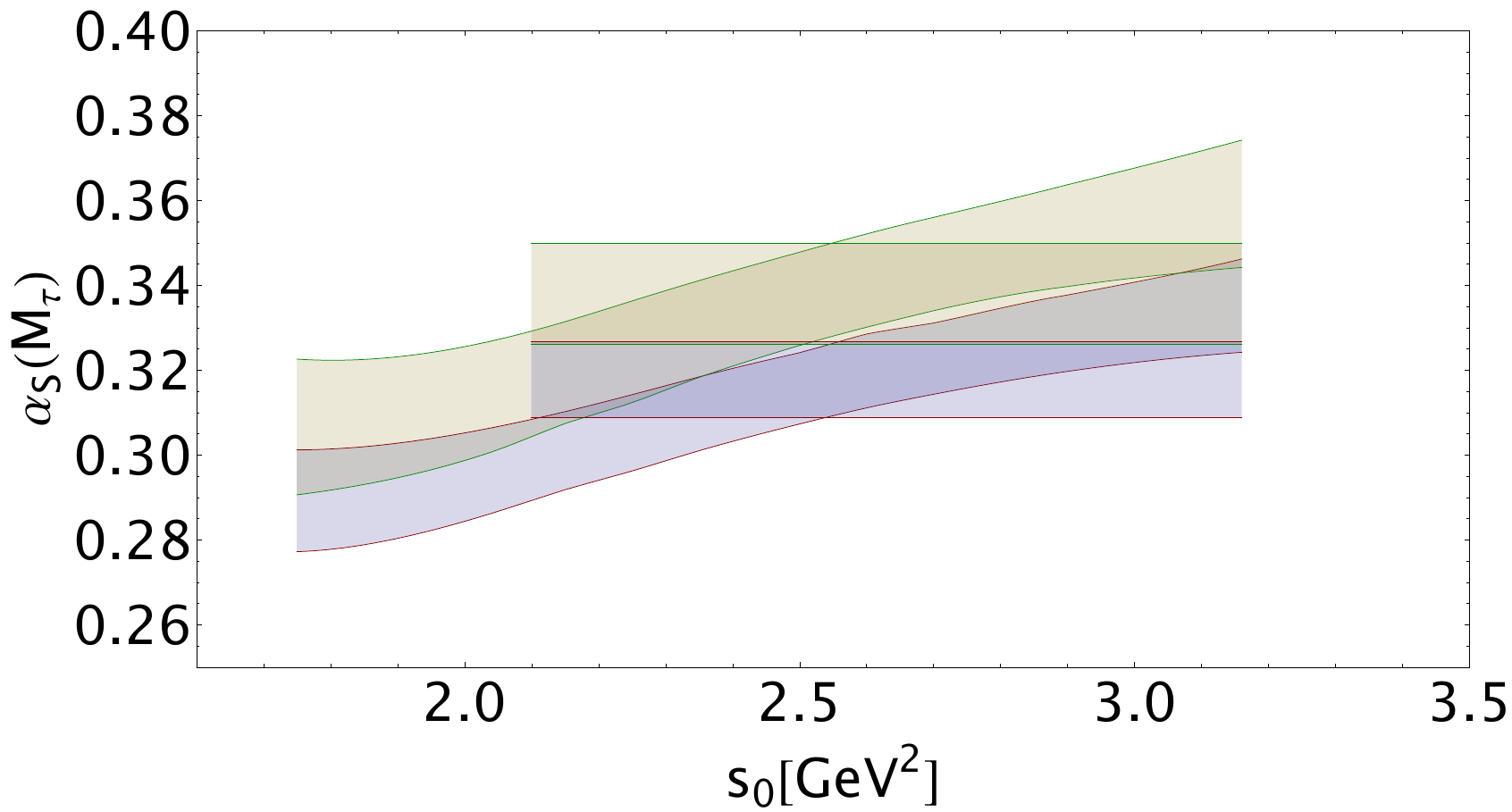}
\caption{\footnotesize  $\alpha_s(M_\tau)$  versus the hypothetical $\tau$-mass squared $s_0$. } \label{fig:as-A}
\end{center}
\vspace*{-0.5cm}
\end{figure} 
\bea
\alpha_s(M_\tau)\vert_A &=&  0.3178(72) (65)~~~~\lrar2~~~~ \alpha_s(M_Z)\vert_A =  0.1182(8)(3)_{evol} ~ ~~~~~   {\rm (FO)} \nnb\\
&=& 0.3380(140)(43) ~~\lrar2~~~~ \alpha_s(M_Z)\vert_A =  0.1206(5)(3)_{evol} ~ ~~~~~   {\rm (CI)}.
\label{eq:as-A}
 \eea
 The 1st error in $\alpha_s(M_\tau)\vert_A$ comes from the fitting procedure. The  2nd one comes from  an estimate of the $\alpha_s^5$ contribution from Ref.\,\cite{SNe}.  At the scale $s_0=$2.5 GeV$^2$ the sum of non-perturbative contributions to the moment normalized to the parton model is:
 \beq
 \delta_{NP,A}\simeq -(7.9\pm 1.1) \times 10^{-2}.
 \eeq
We compare the result with other determinations in the axial-vector channel in Table\,\ref{tab:otherA}
   {\scriptsize
   \begin{center}
\begin{table}[hbt]
\setlength{\tabcolsep}{0.9pc}
  \begin{center}
    {\footnotesize
  \begin{tabular}{ccc cc ll}

&\\
\hline
\oliva Channel  &\oliva$d_6$ &\oliva$-d_8$ &\oliva$\alpha_s(M_\tau)$ FO&\oliva$\alpha_s(M_\tau)$ CI &\oliva Refs.\\
 \hline 
$e^+e^-$&$-(26.3\pm 3.7)$&$-(18.2\pm 0.6)$&0.3081(86)&0.3260(78)& \cite{SNe}\\
\hline
A&$34.4\pm 1.7$&$51.5\pm 2.1$&0.3178(90)&0.3380(140)&This work\,\\
 A&$43.4\pm 13.8$&$59.2\pm 19.7$&$0.3390(180)$&$0.3640(230)$&\,\cite{PICH1}\\
A & $19.7\pm 1.0$&$27\pm 1.2$&{--}&0.3350(120)& \,\cite{ALEPH} \\
A& $9.6\pm 3.3$&$9.0\pm 5.0$&0.3230(160)&0.3470(230) &\,\cite{OPAL} \\

   \hline
\end{tabular}}
 \caption{Values of the QCD condensates from some other $\tau$-moments at Fixed Order (FO) PT series and of $\alpha_s(M_\tau)$ for FO and Contour Improved (CI) PT series.}\label{tab:otherA}.  
 \end{center}
\end{table}
\end{center}
} 

\section{$\alpha_s(M_\tau)$ from the V-A component of $\tau$-decay\label{sec:v-a} }
\d The lowest BNP moment ${\cal R}_{0,V-A}$\,\cite{BNP,BNP2} is expected to be the golden moment for extracting $\alpha_s(M_\tau)$. The reason i s that unlike the V and A moments, the contribution of the leading gluon condensate $\la\alpha_s G^2$ vanishes to lowest order of PT series. 

\d As in previous sections, we fit the data using the same procedure. To calibrate our analysis, we compare the value of our fit at $M_\tau$ with the precise one from ALEPH\,\cite{ALEPH}:
\beq
{\cal R}_{0,V-A}\vert_{\rm our fit}= 3.484 \pm 0.022~~~~ {\cal R}_{0,V-A}\vert_{\rm Aleph} = 3.475\pm 0.011,
\eeq
where our error is conservative as  we have  separately fitted the upper and lower values of the data. 

\d Using as inputs the value of $\la\alpha_s G^2\ra$ in Eq.\,\ref{eq:asg2} and the 
ones\,\cite{SNe2,SNtau24}\,:
\beq
d_{D,V-A} \equiv \frac{1}{2}\ga d_{D,V}+d_{D,A}\dr\,:~~~~~~D=6,8,
\label{eq:d68-VA}
\eeq
where $d_{D,V}\equiv d_D$ in Ref.\,\cite{SNe2} and in this paper, we extract the value of $\alpha_s(M_\tau)$ as a function of $s_0$ (see Fig.\,\ref{fig:as-VA}) from ${\cal R}_{0,V-A}(s_0)$. 
\begin{figure}[hbt]
\begin{center}
\includegraphics[width=10cm]{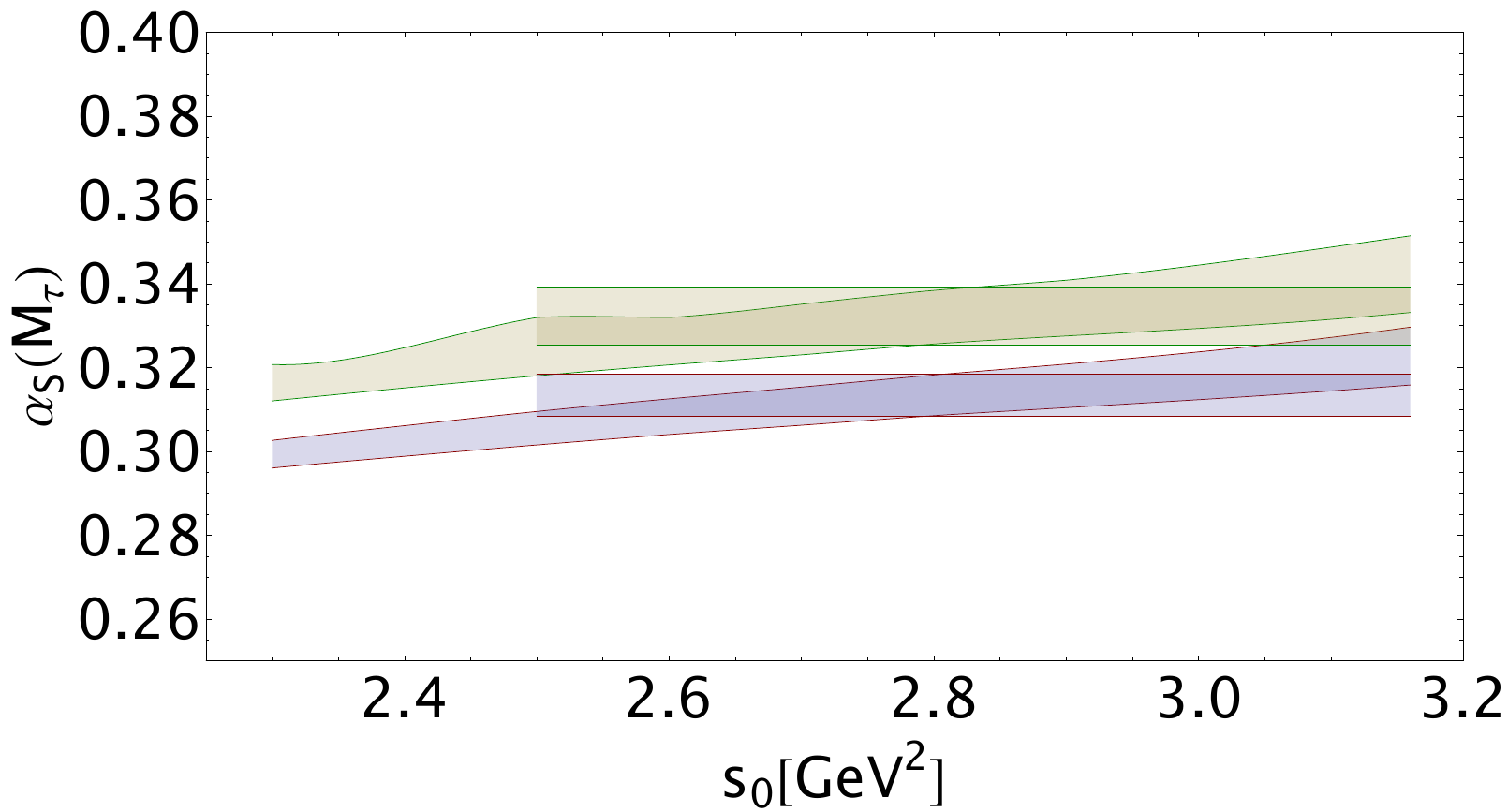}
\caption{\footnotesize  $\alpha_s(M_\tau)$  versus an hypothetical $\tau$-mass squared $s_0$. The upper curves corresponds to CI perturbative series and the lower ones to FO. The horizontal lines come from a least-square fit of the data in the optimal region $s_0\simeq (2.3\sim2.9)$ GeV$^2$.} \label{fig:as-VA}
\end{center}
\vspace*{-0.5cm}
\end{figure} 

\d We notice a stable result in the region $s_0\simeq (2.5\sim 2.6)$ GeV$^2$ though not quite convincing. Then, we consider as a conservative value the one obtained from a least-square fit of the values inside the region  $[2.5,M_\tau^2]$. The optimal result corresponds to $s_0=2.8$ GeV$^2$ (see Fig.\,\ref{fig:as-VA}):
\bea
\alpha_s(M_\tau)\vert_{V-A} &=&  0.3135(51) (65)~\lrar2~ \alpha_s(M_Z)\vert_{V-A} =  0.1177(10)(3)_{evol} ~   {\rm (FO)} \nnb\\
&=& 0.3322(69)(43)~ \lrar2~\alpha_s(M_Z)\vert_{V-A} =  0.1200(9)(3)_{evol} ~    {\rm (CI)}.\nnb\\
\label{eq:as-VA}
 \eea
 
\d The  errors in $\alpha_s(M_\tau)\vert_{V-A}$ come respectively from the fitting procedure and from  an estimate of the $\alpha_s^5$ contribution from Ref.\,\cite{SNe}.  At this scale the sum of non-perturbative contributions to the moment normalized to the parton model is:
 \beq
 \delta_{NP,V-A}\simeq +(2.7\pm 1.1) \times 10^{-4},
 \eeq
 which is completely negligible.
 
\d One can notice from Fig.\,\ref{fig:as-VA} that extracting $\alpha_s(M_\tau)\vert_{V-A} $ at the observed $M_\tau$-mass also tends to overestimate its value\,:
 \beq
 \alpha_s(M_\tau)\vert_{V-A}= 0.3227 (69)(65)~~~{\rm FO},  ~~~~~~~~~~ 0.3423(92)(43)~~~{\rm CI}, 
 \label{eq:as-tau-VA}
 \eeq
 like in the case of the axial-vector channel and $e^+e^-\to$ Hadrons data. 

\section{Average value of $\alpha_s(M_\tau)$ from $e^+e^-$ and $\tau$-decay\label{sec:average} }
Combining the values of $\alpha_s$ from $e^+e^-\to$ hadrons (Eq.\,\ref{eq:as-foci}), the Axial-vector (A) 
(Eq.\,\ref{eq:as-A}) and V-A (Eq.\,\ref{eq:as-VA}) components of $\tau$-decays, we deduce the average from our analysis\,:
\bea
\la \alpha_s(M_\tau)\ra&=&0.3128(51)\, {\rm (FO)} ~ ~~~~~~\lrar2~~~~~~~ \alpha_s(M_Z) = 0.1176(7)_{fit}(3)_{evol.},\nnb\\
 &=& 0.3330 (57) \, {\rm (CI)} ~ ~~~~~~~\lrar2~~~~~~~~~~~~~~~~~=0.1201(7)_{fit}(3)_{evol.}.
 \eea
 The result from the vector component of $\tau$-decays quoted in Table\,\ref{tab:tau} is not included as it has been obtained from the data measured at the physical $\tau$-mass which does not give the optimal result.  The previous mean values can be compared with the PDG average without lattices\,\cite{PDG24}:
 \beq
  \alpha_s(M_Z) = 0.1175(10).
 \eeq
\section*{Declaration of competing interest}  
The author declares that he has no known competing financial interests or personal relationships that could have appeared to influence the work reported in this paper.

\input{bib_qcd25.tex}

\end{document}